\documentclass[12pt]{article}

\usepackage{verbatim, color, amssymb, amsmath,mathtools}

\usepackage[toc,page]{appendix}

\usepackage{amsthm}		
\usepackage{bm}
\usepackage{dsfont}			
\usepackage{multirow}
\usepackage{setspace}
\usepackage[mathscr]{euscript}
\usepackage{fancyhdr}
\usepackage{enumitem}
\usepackage{graphicx}
\usepackage{geometry}
\usepackage{xcolor}
\usepackage{url}
\usepackage{bbm}
\usepackage{tikz}
\usetikzlibrary{positioning, quotes}
\tikzset{RightarrowLine/.style={->,double distance=2pt,thick}}
\usepackage[section]{placeins}

\usepackage{csquotes}
\usepackage[style=ext-authoryear, maxcitenames=1, giveninits=true, maxbibnames=99, sortcites=false, natbib, uniquename=init]{biblatex} 
\DeclareNameAlias{author}{family-given}
\DeclareNameAlias{editor}{family-given}

\renewbibmacro{in:}{%
  \ifentrytype{article}
    {}
    {\bibstring{in}%
     \printunit{\intitlepunct}}}
\DeclareFieldFormat[article]{pages}{#1}

\usepackage{tabularx, booktabs, arydshln}
\newcolumntype{Y}{>{\centering\arraybackslash}X}
\usepackage{lscape}

\usepackage{subfig}
\usepackage{float}

\newtheorem*{Proof*}{Proof}

\def \R {\mathbb{R} }

\def\L{{\cal L}}

\def\F{{\cal F}}

\def\N{{\cal N}}

\def\var{\hbox{var}}
\def\cov{\hbox{cov}}
\def\corr{\hbox{corr}}

\def\EV{\mbox{E}}

\def\Dir{\hbox{Dir}}
\def\DP{\hbox{DP}}

\def\GEM{\hbox{GEM}}

\def\IG{\hbox{Inv-Ga}}

\def\Unif{\hbox{Unif}}

\def\bse{\begin{eqnarray*}}
	\def\ese{\end{eqnarray*}}
\def\be{\begin{eqnarray}}
	\def\ee{\end{eqnarray}}
\def\bq{\begin{equation}}
	\def\eq{\end{equation}}

\def\rank{{\hbox{rank}}}

\def\bone{\bm{1}}

\def\bd{\bm{d}}
\def\bD{\bm{D}}
\def\b1e{\bm{e}}
\def\b1f{\bm{f}}

\def\bM{\bm{M}}
\def\bn{\bm{n}}
\def\bN{\bm{N}}

\def\bP{\bm{P}}

\def\bS{\bm{S}}

\def\bw{\bm{w}}

\def\bx{\bm{x}}
\def\bX{\bm{X}}
\def\by{\bm{y}}

\def\tw{\widetilde{w}}
\def\tbw{\widetilde{\bm{w}}}
\def\ty{\widetilde{y}}
\def\yd{y^{\dagger}}

\def\tmu{\widetilde{\mu}}

\def\tbeta{\widetilde{\beta}}
\def\teta{\widetilde{\eta}}
\def\txi{\widetilde{\xi}}

\def\tF{\widetilde{F}}
\def\tM{\widetilde{M}}
\def\tbM{\bm{\widetilde{M}}}

\def\tS{\widetilde{S}}
\def\tbS{\bm{\widetilde{S}}}

\newcommand{\FF}{\mathcal{F}}

\def\hP{\widehat{P}}
\def\bhP{\bm{\widehat{P}}}
\def\hR{\widehat{R}}

\newcommand{\bmu}{\bm{\mu}}

\newcommand{\bpi}{\bm{\pi}}

\newcommand{\btheta}{\bm{\theta}}
\newcommand{\bth}{\bm{\theta}}

\newcommand{\bSigma}{\bm{\Sigma}}

\def\Rstar{R^{\star}}

\def\Psis{\Psi^{\star}}

\renewcommand{\ij}{^{(i,j)}}
\newcommand{\Pij}{P^{(i,j)}}
\newcommand{\tht}{\widetilde\theta}

\newcommand{\sigst}{\widetilde{\sigma}^{2}}
\newcommand{\mut}{\widetilde\mu}
\newcommand{\betat}{\widetilde\beta}

\def\simind{\stackrel{\mbox{\scriptsize{ind}}}{\sim}}
\def\simiid{\stackrel{\mbox{\scriptsize{iid}}}{\sim}}
\def\eqd{\stackrel{\mbox{\scriptsize{d}}}{=}}

\def\d{\mathrm{d}}

\renewcommand\footnoterule{\kern-3pt \hrule \textwidth 2in \kern 2.6pt}

\def\boxit#1{\vbox{\hrule\hbox{\vrule\kern6pt \vbox{\kern6pt \textcolor{blue}{#1}\kern6pt}\kern6pt\vrule}\hrule}}

\def\authorfootnote#1{{\let\thefootnote\relax\footnotetext{#1}}}

\allowdisplaybreaks

\definecolor{grau}{rgb}{0.55,0.55,0.55}

\begin{document}
\newrefsection

\begin{center}
{\LARGE{
\bf	Separate Exchangeability as Modeling Principle in Bayesian Nonparametrics}}
	\baselineskip=12pt
	\vskip 24pt 
\end{center}

\begin{center}
	\small
	\baselineskip=14pt
    Giovanni Rebaudo$^{a}$ (giovanni.rebaudo@unito.it) \\
    Qiaohui Lin$^{b}$ (qiaohui.lin@utexas.edu)\\
    Peter M\"uller$^{c}$ (pmueller@math.utexas.edu) \\
    \vskip 3mm
    $^{a}$ESOMAS Dept., University of Torino and Collegio Carlo Alberto, IT\\
    \vskip 4pt 
    $^{b}$Genentech, USA
    \vskip 4pt 
    $^{c}$SDS \& Mathematics Depts., University of Texas at Austin, USA\\ 
\end{center}
	
\vskip 24pt 
	
\begin{center}
	{\Large{\bf Abstract}} 
\end{center}
\baselineskip=16pt
We argue for the use of separate exchangeability as a modeling principle in Bayesian nonparametric (BNP) inference.
Separate exchangeability is \emph{de facto} widely applied in the Bayesian parametric case, e.g., it naturally arises in simple mixed models.
However, while in some areas, such as random graphs, separate and (closely related) joint exchangeable models are widely used, they are curiously underused for several other applications in BNP.
We briefly review the definition of separate exchangeability, focusing on the implications of such a definition in Bayesian modeling.
We then discuss two tractable classes of models that implement separate exchangeability, which are the natural counterparts of familiar partially exchangeable BNP models.

The first is nested random partitions for a data matrix, defining a partition of columns and nested partitions of rows, nested within column clusters.
Many recent models for nested partitions implement partially exchangeable models related to variations of the well-known nested Dirichlet process.
We argue that inference under such models in some cases ignores important features of the experimental setup.
We obtain the separately exchangeable counterpart of such partially exchangeable partition structures.

The second class is about setting up separately exchangeable priors for a nonparametric regression model when multiple sets of experimental units are involved.
We highlight how a Dirichlet process mixture of linear models, known as ANOVA DDP, can naturally implement separate exchangeability in such regression problems.
Finally, we illustrate how to perform inference under such models in two real data examples.
\vskip 10pt 

\par\medskip\noindent\underline{\bf Short/Running Title}: Separate Exchangeability in BNP.
\par\medskip\noindent\underline{\bf Corresponding Author}: Giovanni Rebaudo (giovanni.rebaudo@unito.it).

\vskip 10mm	
\newlength{\gnat}
\setlength{\gnat}{20pt}
\baselineskip=\gnat

\section{Introduction} \label{sec: intro}
We argue for using separate exchangeability as a modeling principle and a unifying framework for data that involves multiple sets of experimental units, such as columns and rows in a data matrix.
Under Bayesian parametric inference, separate exchangeability is \emph{de facto} often naturally preserved, e.g., by introducing additive Gaussian row- or column-specific random effects in mixed effects models.

While exchangeability and partial exchangeability have proven to be powerful principles for statistical modeling \citep[see e.g.,][]{bernardo2009bayesian}, separate exchangeability has been curiously underused as a modeling principle in some of the Bayesian nonparametric (BNP) literature.
We introduce and discuss general tractable classes of models that are natural counterparts or variations of widely used partially exchangeable BNP models, allowing us to perform flexible BNP inference under separate exchangeability in practice.

First, we review the definition of separate exchangeability, focusing on the implications in Bayesian modeling and highlighting differences from other homogeneity assumptions that are expressed by way of different probabilistic invariances.
The discussion is not meant to be a detailed probabilistic account of different notions of exchangeability \citep[see e.g.,][]{aldous1985exchangeability,kallenberg2005probabilistic}.
Rather, we focus on the implications of separate exchangeability in Bayesian data analysis, including probabilistic dependence and the implied borrowing of strength in the learning process.
See also \cite{orbanz2014bayesian} and \cite{foti2015survey} for insightful reviews of the topic focused on representation theorems and BNP models.
Separate exchangeability facilitates inference under homogeneity assumptions that maintain the identity of experimental units in situations with the same experimental units measured under another type of experimental unit, such as for array/matrix data.

In particular, if the data (or the design matrix) is a rectangular array, the nature of rows and columns as different experimental units is preserved.
Such experimental data formats are ubiquitous across many fields.
Quoting \cite{nguyen2017bayesian}: ``\emph{Multivariate, biological data is usually represented in the form of matrices, where features (e.g.\ genes, species) represent one dimension and observations (e.g.\ samples, cells) the other.''} 

To further motivate this work, we rephrase and generalize important ad hoc arguments in the BNP literature against the use of dependent random partitions that arise from ties under (conditionally) independent and identically distributed (i.i.d.) sampling from dependent discrete random probabilities \citep{lee2013nonparametric, page2022dependent}.
We characterize the same arguments in terms of subjective probabilistic invariance assumptions as a modeling principle.
On one hand, this allows us to gain a novel understanding of these arguments; on the other hand, this allows data analysts to assess when such assumptions are appropriate (e.g., coherent with their judgment and the experimental design) and when other probabilistic invariances are more effective.

We provide general guidance to define and apply tractable separately exchangeable models as a natural alternative and counterpart of widely used partially exchangeable BNP models.
This allows us to preserve the flexibility of existing BNP models while respecting the experimental design in prior elicitation and the learning process, for example, in the context of matrix data.
More precisely, we introduce and study in this unifying framework two widely applicable classes of models that implement inference under separate exchangeability, which are the natural counterparts of familiar partially exchangeable models.
We also illustrate how to perform inference under such models in the context of two typical real data applications.

The first class of separately exchangeable models allows defining separately exchangeable partition structures via nested partitions, generalizing \cite{lee2013nonparametric}.
We illustrate how to perform inference under such a model on microbiome data similar to what is done in a flexible BNP analysis by \cite{denti2023common}, but also accommodating the data design by implementing a separately exchangeable model.

The second class of models makes a similar point with BNP regression models by considering widely used dependent Dirichlet process (DDP) models and highlighting how these can easily be modified to accommodate separately exchangeable structures.
We illustrate the model with an analysis of protein time course data collected on a set of proteins and a set of subjects.

The rest of this article is organized as follows.
Section \ref{sec: exc} reviews different notions of exchangeability and how they relate to modeling and borrowing of information in Bayesian inference.
Section \ref{sec: BNP arg} rephrases and generalizes two important arguments in the BNP literature against the use of random partition models that arise from dependent discrete random measures (for rectangular array data), providing novel understanding and practical guidance.
Sections \ref{sec: BNP RCE} and \ref{sec: sep ANOVA} present and discuss general tractable classes of separately exchangeable BNP models that preserve the identity of the types of experimental units in rectangular data arrays.
Specifically, Section \ref{sec: BNP RCE} considers nested clustering and Section \ref{sec: sep ANOVA} considers BNP regression.
Section \ref{sec: illu} illustrates how to perform inference under the earlier introduced exchangeable models on two examples.
Section \ref{sec: discussion} contains concluding remarks and suggests future work.
Substantial additional details, including Markov chain Monte Carlo (MCMC) based posterior inference algorithms, are presented in the supplementary materials.

\section{Exchangeability as a Modeling Principle}
\label{sec: exc}
To perform inference and prediction, we rely on some notion of homogeneity across observations that allows us to leverage information from the sample $x_{1:n} =(x_{1},\ldots,x_{n})$ to deduce inference about future observations $x_{n+1:n+m}$.
\cite{finetti1937prevision} refers to this as ``\emph{analogy}''. 
Importantly, such an assumption allows performing predictions based on induction that would not be justified without some subjective homogeneity assumptions.
See Hume's discussion of the ``\emph{problem of induction}'' in \cite[Book 1, Part 3, Section 6]{hume1739treatise}.
In Bayesian statistics, the assumptions are stated in the language of probability, and learning is performed via conditional probability.
Throughout, we use the same lowercase symbols for random variables and realizations, only to simplify notation.

\subsection{Exchangeability and Partial Exchangeability}
\paragraph*{Exchangeability.}
A fundamental assumption that allows such generalization in Bayesian learning is exchangeability, that is, invariance of the joint law of the data with respect to permutations of the observation indices, implying that the order of the observations is irrelevant in the learning process and one can deduce inference for $x_{n+1:n+m}$ from observations $x_{1:n}$.
More precisely, a sequence $x_{1:n}$ is judged exchangeable if 
\begin{equation}\label{eq:def_Fin_Exc}
	x_{1:n} \eqd x_{\pi(1:n)}
\end{equation}
for any permutation $\pi$ of $[n] \coloneqq \{1,\ldots,n\}$.
Here, $\eqd$ denotes equality in distribution.
If the observable $x_{1},\ldots,x_{n}$ can be represented as a sample from an infinite exchangeable sequence $(x_{i})_{i\ge 1}$, that is finite exchangeability holds for any sample size $n \ge 1$, we say that $x_{1},\ldots,x_{n}$ is an extendable exchangeable sequence.
De Finetti's theorem \citep{finetti1930funzione, finetti1937prevision, hewitt1955symmetric, kingman1978uses} states that a sequence $x_{1}, \ldots, x_{n}$ is extendable exchangeable if and only if it
can be expressed as conditionally i.i.d.\ from a probability measure $P$
\begin{equation}
    x_{i} \mid P \simiid P, \quad i=1,2,\ldots \quad \quad 
    P \sim \L.\label{eq:deFExc}
\end{equation}
Note that $x_{1},\ldots,x_{n}$ is an exchangeable extendable sequence if it can be seen as a projection of an infinite exchangeable sequence $(x_{i})_{i \ge 1}$ on the first $n$ coordinates.
The characterization of infinite exchangeability as a mixture of i.i.d.\ sequences highlights the fact that the homogeneity assumption of (infinite) exchangeability in Bayesian learning is equivalent to assuming an i.i.d.\ sequence in the frequentist paradigm.
The de Finetti measure $\L$ can be interpreted as the prior in the Bayes-Laplace paradigm.
If $P$ is restricted to a parametric family, we can write $P$ as $P_{\theta}$, where $\theta$ denotes a finite-dimensional random parameter and $\L$ reduces to a prior probability model on
$\theta$. 
However, if $P$ is unrestricted then $\L$ takes the form of a BNP prior for $P$
\citep[see e.g.,][]{ghosal2017fundamentals}.

Note that the unknown probability $P$ arises from an assumption on observable random variables, justifying inference on the parameters/latent quantities also in terms of measurable quantities \citep[see e.g.,][]{regazzini1997finetti}.
This is closely related to a predictive approach to statistics that is becoming increasingly popular in statistics and machine learning.
See e.g., \cite{fortini2000exchangeability, fortini2012predictive, fortini2016predictive, fortini2025exchangeability, berti2025probabilistic} for interesting discussions of the predictive approach and characterization results for the prior probability measure in terms of predictive sequences under exchangeability and partial exchangeability.

\paragraph*{Partial exchangeability.}
In real-world applications, the assumption of exchangeability is often too restrictive.
To quote \cite{finetti1938condition}: 
\emph{``But the case of exchangeability can only be considered as a limiting case: the case in which this ‘analogy’ is, in a certain sense, absolute for all events under consideration.
[\ldots]
To get from the case of exchangeability to other cases which are more general but still tractable, we must take up the case where we still encounter ‘analogies’ among the events under consideration, but without attaining the limiting case of exchangeability.''
} 
Moreover, depending on the design of the experiment, it is often meaningful to generalize exchangeability to less restrictive invariance assumptions that allow us to introduce more structure into the prior, and thus into the learning mechanism.

If the data are collected in different, related populations, a simple generalization of exchangeability is partial exchangeability.
Let $\bX=(x_{i,j}: i=1,\ldots,I_{j},\, j=1,\ldots,J)$ denote a data array, where $j$ is the label of the population from which $x_{i,j}$ is collected.
We say that $\bX$ is partially exchangeable if the joint law is invariant under different permutations of the observations within each population 
\begin{align} \label{eq: PartExc}
\begin{split}
    &(x_{i,j}: i=1,\ldots,I_{j}, \, j=1,\ldots,J)\\
    &\quad \eqd (x_{\pi_{j}(i),j}: i=1,\ldots,I_{j}, \, j=1,\ldots,J),
\end{split}
\end{align}
for any choice of finite permutations $\pi_{1},\ldots,\pi_{J}$ of $[I_{1}]$, $\ldots$, $[I_{J}]$.
Partial exchangeability entails that the order of the observations is irrelevant in the learning mechanism, up to preserving the information of the population memberships.
If partial exchangeability holds for any sample sizes $(I_{1},\ldots, I_{J}) \in \mathbb{N}^{J}$ the ``\emph{analogy}'' assumption of partial exchangeability can be characterized in terms of latent quantities (e.g., parameters) via de Finetti's theorem \citep{finetti1938condition, aldous1985exchangeability},
\begin{equation}
\begin{split}
    &x_{i,j} \mid (P_{1},\ldots,P_{J}) \simind P_{j} \quad j=1,\ldots,J, \, i=1,\ldots\\
    &(P_{1},\ldots,P_{J}) \sim \L.
\end{split}
\label{eq: deFPart}
\end{equation}
Therefore, partially exchangeable extendable arrays can be thought of as decomposable into different conditionally independent exchangeable populations.
As in \eqref{eq:deFExc}, the characterization in \eqref{eq: deFPart} does not restrict the distributions associated with the different populations to any parametric family.
The $P_{j}$'s would usually be assumed to be dependent, allowing borrowing of strength across blocks.

Note that exchangeability is a degenerate special case of partial exchangeability, which corresponds to ignoring the information on the specific populations $j$ from which the data are collected, i.e., ignoring the known heterogeneity.
The opposite, also degenerate, extreme case corresponds to modeling data from each population independently, i.e., ignoring similarities between populations.
See \cite{aldous1985exchangeability,kallenberg2005probabilistic,austin2008exchangeable,aldous2010uses} for detailed probabilistic accounts on different exchangeability assumptions and \cite{orbanz2014bayesian} or \cite{foti2015survey} for discussions in the context of BNP models.

The way partial exchangeability preserves heterogeneity is easiest to see by considering correlations between pairs of observations.
As desired, partial exchangeability allows increased dependence between two observations arising under the same experimental condition, compared to the dependence between observations arising from different populations.
For instance, under \eqref{eq: PartExc} it is possible that
\begin{equation}
    \corr(x_{i,j},x_{i^{\prime}, j}) >
    \corr(x_{i,j},x_{i^{\prime}, j^{\prime}}),
    ~~~
    j \ne j^{\prime},\, i \ne i^{\prime}
    \label{eq: corrPE}
\end{equation}
while by definition of exchangeability, under \eqref{eq:def_Fin_Exc}
\begin{equation} \label{eq:corrPE2}
    \corr(x_{i,j},x_{i^{\prime}, j}) = \corr(x_{i,j},x_{i^{\prime}, j^{\prime}}),
\end{equation}
entailing that in the learning process we can borrow information across groups, but to learn about a specific population $j$, the observations recorded in the same population $j$ are more informative than observations from another population $j^{\prime}$.
Here and in the following, we assume that $x_{i,j}$ are real-valued, square-integrable random variables such that the stated correlations are well-defined.
Moreover, for an extendable partial exchangeable array such that
within column correlations are equal, i.e., $\corr(x_{i,j},
x_{i^{\prime}, j})=\corr(x_{i,j^{\prime}}, x_{i^{\prime}, j^{\prime}})$, we have
\[
\corr(x_{i,j},x_{i^{\prime}, j}) \ge \corr(x_{i,j},x_{i^{\prime}, j^{\prime}}).
\]
In words, the correlation between observations (that can measure some borrowing of information under Bayesian inference) within a column is greater than or equal to the one across columns.
See Section S.2 of the supplementary materials for a simple proof.
Going beyond correlations between individual observations, one can see a similar pattern for groups of observations under populations $j$ and $j^{\prime}$.
See Section \ref{sec: SepExc} for more discussion.

\paragraph*{Statistical inference.}
From a statistical inference perspective, partial exchangeability is a framework that allows borrowing of information across related populations while preserving population heterogeneity. 
Partially exchangeable models are widely used in many applications, including meta-analysis, topic modeling, and survival analysis when the design matrix records different experimental units (e.g.\ patients) under related experimental conditions (e.g.\ hospitals).

Flexible learning can be achieved assuming dependent non-parametric priors for the vector of random probabilities $P_{j}$ in \eqref{eq: deFPart}.
An early proposal appeared in \cite{cifarelli1978problemi}, but the concept was widely taken up in the literature only after the seminal paper of \cite{maceachern1999dependent, maceachern2000dependent} introduced the DDP as a prior over a family of random probability measures $\F=\{F_{x}: x \in X\}$, where the random probability measures $F_{x}$ are indexed by covariates $x$, an instance of which could be used for $\L$ in \eqref{eq: deFPart}.
See \cite{quintana2022dependent} for a recent review of DDP models.

Finally, in anticipation of the later examples, we note that while in a stylized setup it is convenient to think of $x_{i,j}$ in \eqref{eq: deFPart} as the observable data, in many applications, the discussed symmetry assumptions are used to construct prior probability models for (latent) parameters or data. 
The $x_{i,j}$ might be, for example, cluster membership indicators for bi-clustering models, or any other latent quantities in a larger hierarchical model.
The same comment applies to the upcoming discussion of separate exchangeability.
 
\subsection{Separate Exchangeability}\label{sec: SepExc}
Similar to exchangeability being too restrictive when observations are arranged in different populations, partial exchangeability can be too restrictive when the same experimental unit is recorded across different blocks of observations (e.g., the same patient is recorded in various hospitals). 
Here and elsewhere, by restrictive assumption, we mean an unnatural judgment that a reasonable investigator would not choose to adopt since it discards structural information about the design matrix.
A similar case arises when two types of experimental units are involved, and observations are recorded for combinations of the two types of units, as is common in many experimental designs (e.g., in the first motivating example in Section \ref{sec: illu} the same type of microbiome is observed across different subjects).
 
In such a case, a simple but effective homogeneity assumption that preserves the information on the experimental design is separate exchangeability, that is, invariance of the joint law under different permutations of indices related to the two types of units (or blocks).
If data is arranged in a data matrix with rows and columns corresponding to two different types of units (or blocks), this reduces to invariance with respect to arbitrary permutations of row and column indices.
More precisely, a data matrix is separately exchangeable if
\begin{equation*}
    x_{1:I,1:J} \eqd x_{\pi_{1}(1:I),\pi_{2}(1:J)}
\end{equation*}
for separate permutations $\pi_{1}$ and $\pi_{2}$ of rows and columns, respectively.

The notion of separate exchangeability formally reflects the known design in the learning mechanism.
That is, it introduces more dependence between two values recorded from the same experimental unit than between values recorded in different experimental units.
Importantly, this entails borrowing more information across measurements obtained from the same experimental unit (in the earlier case) than from two different units (in the latter case).
Note also that partial exchangeability of $x_{1:I, 1:J}$ grouped w.r.t.\ columns plus exchangeability of the columns (which is naturally assumed when there is no natural ordering between columns) is a stronger homogeneity judgment
than separate exchangeability in a similar way as exchangeability is an extreme case of partial exchangeability (i.e.\ we lose some structure).
Figure \ref{fig:exch} summarizes the relations between such homogeneity assumptions.

Similarly to \eqref{eq: corrPE} and \eqref{eq:corrPE2}, under separate exchangeability it is possible that
\begin{align}
    \corr(x_{i,j}, x_{i,j^{\prime}}) > \corr(x_{i,j},x_{i^{\prime}, j^{\prime}}), ~~~ j \ne j^{\prime},\, i \ne i^{\prime}
\label{eq:corrSE}
\end{align}
while by the definition of partial exchangeability, under
\eqref{eq: PartExc} we always have
\begin{equation*}
\corr(x_{i,j},x_{i,j^{\prime}}) = \corr(x_{i,j},x_{i^{\prime}, j^{\prime}}).
\end{equation*}
More precisely, inequality \eqref{eq:corrSE} is always true with $\ge$ for an extendable separately exchangeable array (and it can also be true with $<$ for a non-extendable separately exchangeable array) and there exists a separately exchangeable array such that the strict inequality in \eqref{eq:corrSE} holds.
See Proposition 15.2 in \cite{aldous1985exchangeability} for the proof and below for a more detailed discussion on the borrowing of information under partial and separate exchangeability.

\begin{figure}
\centering
\begin{tikzpicture}[
    node distance = 12mm and 6mm,
       box/.style = {rectangle, draw, fill=#1, 
                     minimum width=12mm, minimum height=7mm}
                        ]
\node (n1) [box=blue!10,align=center] {Exchangeability: 
$x_{1:I,1:J} \eqd x_{\pi(1:I,1:J)}$};
\node (n2) [box=white,below= 1cm of n1,align=center] {Partial Exchangeability: 
$x_{1:I,1:J} \eqd (x_{\pi_{j}(1:I),j})$ \\
+
\\
Exchangeability of Columns: $x_{1:I,1:J} \eqd x_{1:I,\pi(1:J)}$};
\node (n3) [box=blue!10,below left= 1cm and -3cm of n2, align=center]
{Partial Exchangeability:\\
$x_{1:I,1:J} \eqd (x_{\pi_{j}(1:I),j})$};
\node (n4) [box=blue!10,below right=1cm and -3cm of n2, align=center] 
{Separate Exchangeability:\\
$x_{1:I,1:J} \eqd x_{\pi_{1}(1:I),\pi_{2}(1:J)}$};
%
\draw[RightarrowLine] (n2) -- (n1);
\draw[RightarrowLine] (n3) -- (n2);
\draw[RightarrowLine] (n4) -- (n2);
\end{tikzpicture}
\caption{Relations between probabilistic invariances (i.e., homogeneity assumptions) on a matrix $x$, where we use the notations $x=x_{1:I,1:J} = (x_{1:I,j})$.
An arrow indicates that a target is a special case of the origin of the arrow.
Being a special case entails stronger invariance and thus less structure.} 
\label{fig:exch}
\end{figure}

Finally, we conclude this section by reviewing a version of de Finetti's theorem for separately exchangeable arrays.
If $x_{1:I,1:J}$ is extendable, that is, it can be seen as a projection of $(x_{i,j}: i=1,\ldots, \, j=1,\ldots)$, a representation theorem in terms of latent quantities was proven independently by \cite{aldous1981representations} and \cite{hoover1979relations}.
See also \cite{kallenberg1989representation} and references therein.
More precisely, an extendable matrix $x_{1:I,1:J}$ is separately exchangeable if and only if
\begin{equation}\label{th: AldHoov}
    x_{i,j} = f(\theta,\xi_{i}, \eta_{j}, \zeta_{i,j}),
\end{equation}
for some measurable function $f:[0,1]^4 \rightarrow \mathbb{R}$ and i.i.d.\ $\Unif(0,1)$ random variables $\theta, \xi_{i}, \eta_{j}$ and $\zeta_{i,j}$, $i , j \in \mathbb{N}$.
The representation theorem in \eqref{th: AldHoov} implies at first sight less strict, but equivalent representation theorem in which the uniform distributions are replaced by any distributions $p_{\xi}$, $p_{\eta}$, and $p_{\zeta}$, as long as independence is maintained (together with a corresponding change of the domain for $f$).
That is,
\begin{align}\label{th: AldHoovA}
\begin{split}
  &p(x_{1:I,1:J}) =
  \int p_{\theta}(\theta)
  \prod_{i=1}^{I} p_{\xi}(\xi_{i})\prod_{j=1}^{J} p_{\eta}(\eta_{j})\\
  &\quad \prod_{i,j}p(x_{i,j} \mid \theta,\xi_{i},\eta_{j})\,
  d\theta \, d\xi_{1:I} \, d\eta_{1:J},
\end{split}
\end{align}
as stated in \cite{roy2009mondrian}.
Similar to \eqref{eq:deFExc} or \eqref{eq: deFPart} model \eqref{th: AldHoovA} can be stated as a hierarchical model 
\begin{align} 
\begin{split}
  &x_{i,j}\mid \theta, \eta_{j}, \xi_{i} \simind P_{\theta,\xi_{i},\eta_{j}}, \quad i=1,\ldots, \, j=1,\ldots\\
  &\theta \sim p_{\theta} \, \perp \, \xi_{i} \simiid p_{\xi} \, \perp \, \eta_{j} \simiid p_{\eta},\, i=1,\ldots,\, j=1,\ldots
\end{split}\label{eq: sepC}
\end{align}
where $P_{\theta,\xi_{i},\eta_{j}}$ is the law of $p(x_{i,j} \mid \theta, \xi_{i},\eta_{j})$ in \eqref{th: AldHoovA}.
Note that the randomness implied by $\zeta_{i,j}$ in \eqref{th: AldHoov} is incorporated in the random sampling for $x_{i,j}$ mechanism in \eqref{th: AldHoovA} and \eqref{eq: sepC}. 
Finally, in both \eqref{th: AldHoovA} and \eqref{eq: sepC}, the probability models can be additionally indexed with unknown hyperparameters.

We discuss in more detail the nature of borrowing of information under partial and separate exchangeability.
For easier exposition, when it is sufficient to illustrate the idea, we focus on a matrix $\bX=[x_{i,j}, j=1,\ldots, J,\, i=1, \ldots, I]$, with two columns, i.e.\ $J=2$.
Both partial and separate exchangeability imply marginal exchangeability within columns.
Moreover, we exclude the degenerate cases of (marginal) independence between $x_{1:I,j}$ and $x_{1:I,j^{\prime}}$, when no borrowing of strength occurs under Bayesian learning, as well as the fully exchangeable case when we lose heterogeneity across columns.

Earlier we discussed how, in contrast to partial exchangeability, under separate exchangeability one can have increased dependence between $x_{i,1}$ and $x_{i,2}$ compared to dependence between $x_{i,1}$ and $x_{i^{\prime},2}$.
\footnote{Note that sometimes it might be desirable to introduce negative dependence (repulsion), e.g.\ as $\mbox{corr}(x_{i,1},x_{i^{\prime},2})<0$.} 
We now extend the discussion to the borrowing of information also on other functionals of interest of $x_{1:I,1}$ and $x_{1:I,2}$ focusing on clustering, potentially even while assuming $x_{i,1}$ and $x_{i^{\prime},2}$ independent for any $i, i^{\prime} \in [I]$, but $x_{1:I,1}$ and $x_{1:I,2}$ dependent.
An example can be obtained by introducing dependent partitions, in terms of marginal frequencies and number of clusters, but independent parameters that index cluster-specific sampling models, independent across columns.
In the following, we focus on partitions of $[I]$ and assume discrete $P_{\theta,\xi,\eta}$.
In that case, ties of $(x_{i,j},\, i=1,\ldots)$ define a partition of $[I]$ for each column $j$ (and similarly for rows, but for simplicity we only focus on columns).
Let $\Psi_{j}$ denote this partition in column $j$.

A common context where such random partition models arise is when $x_{i,j}$ are latent indicators in a mixture model for observable data $y_{i,j}$, with a top-level sampling model $p(y_{i,j} \mid x_{i,j}=\ell, \varphi_{\ell})$, where $\varphi_{\ell}$ are additional cluster-specific parameters.
\footnote{Note that the case of borrowing of information across $j$ via the partition with pairwise independence across columns $j$ can be obtained by introducing column-specific cluster parameters: $p(y_{i,j} \mid x_{i,j}=\ell, \varphi_{j, \ell})$ and assuming $\varphi_{j, \ell}$ to be independent across columns $j$.}

Similarly, a partially exchangeable model \eqref{eq: deFPart} with discrete probability measures $P_{j}$ defines a random partition $\Psi_{j}$ in column $j$.
If we then induce dependence between $\Psi_{1}$ and $\Psi_{2}$ (by way of dependent $P_{j}$), it is possible to borrow information about the law of $\Psi_{j}$, for example, the distribution of the number of clusters.
See \cite{franzolini2025multivariate} for definition and probabilistic characterizations of partially exchangeable partitions.
However, by definition of partial exchangeability \eqref{eq: PartExc}, with only partial exchangeability, it is not possible to borrow information about the actual realizations $\Psi_{1}, \Psi_{2}$.
For example, under a partially exchangeable partition
\begin{align}\label{eq: PEclust}
\begin{split}
    &p(\{x_{1,j^{\prime}}= x_{2,j^{\prime}}\} \mid \{x_{1,j}= x_{2,j}\})\\
    &\quad= p(\{x_{1,j^{\prime}}= x_{3,j^{\prime}}\} \mid \{x_{1,j}= x_{2,j}\}).
\end{split}
\end{align}
In contrast, under separate exchangeability, it is possible to have
\begin{align}\label{eq: SEclust}
\begin{split}
    &p(\{x_{1,j^{\prime}}= x_{2,j^{\prime}}\} \mid \{x_{1,j}= x_{2,j}\})\\ 
    & \quad > p(\{x_{1,j^{\prime}} = x_{3,j^{\prime}}\} \mid \{x_{1,j}= x_{2,j}\}).
\end{split}
\end{align}
From a statistical perspective, this means that under separate exchangeability the fact that two observations are clustered together (e.g.\ $1$ and $2$) in one column can increase the probability that the same two observations are clustered together in another column without increasing, in the same way, the probability that one of the two observations (e.g.\ $1$) is clustered together with different observations (e.g.\ $3$).
This difference in the probabilistic structures, and thus in the borrowing of information under Bayesian learning, is particularly relevant when observations indexed by $i$ have a meaningful identity in a particular application.
For example, in one of the examples in Section \ref{sec: Sep RPM}, units $i=1,2,3$ refer to three different types of microbiomes (OTU), and columns $j=1,2$ refer to two subjects.
In this case, \eqref{eq: SEclust} implies that seeing OTUs 1 and 2 clustered together in subject $j$ can increase the probability of seeing the same OTUs co-cluster in subject $j^{\prime}$.
Importantly, the same is not possible under partial exchangeability without increasing in the same way also the probability of clustering OTUs 1 and 3, by \eqref{eq: PEclust}, even if OTU 3 has a completely different biological meaning than OTU 2.
In the actual application, $x_{i,j}$ will refer to parameters in a hierarchical model.

In Section \ref{sec: Sep RPM}, we discuss an effective way to define flexible, but analytically and computationally tractable, non-degenerate separately exchangeable partitions.
We first cluster columns and then set up nested partitions of the rows, nested within column clusters.
That is, all columns in the same column cluster share the same nested partition of rows.

\subsection{DDP models and exchangeability}
\label{sec: BNP arg}
In the BNP literature, we can find important ad hoc arguments against the use of random partition models that arise from dependent discrete random probabilities such as the DDP \citep[see e.g.,][]{quintana2022dependent}.
We note that such random partition models are partially exchangeable.
Here, we rephrase such arguments in terms of subjective probabilistic invariance assumptions, i.e., modeling principles, allowing us to gain a novel understanding of these arguments and assess when such assumptions are appropriate (e.g., coherent with expert judgment or the design of their experiment) and when other probabilistic invariances are more effective.

A related argument appears in Section 1.4 ``\emph{Current Approaches and Limitations}'' in \cite{lee2013nonparametric}.
The authors introduced an effective nested separately exchangeable partition model to perform nested clustering of matrix data where the columns denote proteins and the rows represent the experimental units (``\emph{samples}'').
They highlight the limitation of the existing nested partition models, such as the ones arising from \cite{rodriguez2008nested, wade2011enriched}, and comment in the context of modeling a data matrix that such models allow sharing the prior distribution of the partition of the samples between columns, but not on their actual
realizations as wished in their application. 
More recently, \cite{page2022dependent} introduced an interesting dependent random partition model motivated by a longitudinal data analysis where the data are represented as a matrix with columns and rows denoting times and observations, respectively.
They argue that modeling longitudinal clustering via the dependent random partitions arising from dependent discrete priors, as commonly done in the BNP literature, is problematic from a modeling and applied perspective.

We agree with both arguments in the context of the specific data matrices in which the experimental units (sample/observation) are shared across the other dimension (protein/time).
Although the two articles do not explicitly link their arguments to general principles of invariance, their arguments can be seen as special cases of our argument that partially exchangeable random partition models can be inappropriate when the same experimental unit is recorded in multiple groups.
Understanding the critiques as related to a more general probabilistic invariance principle also highlights that the same concerns would not apply to data arrays where the experimental units are different in different groups (e.g., different patients in different hospitals).

\section{Nested Partitions}\label{sec: BNP RCE}
\subsection{Separate vs Partial Exchangeability in Nested Partitions} 
\label{sec: Sep RPM}
We consider a common data structure with a matrix of (continuous) measurements $y_{i,j}$.
We assume real-valued data, but note that the discussion can be generalized for observations taking values in an arbitrary Polish space.
Assume then that one type of experimental unit $i=1,\ldots, I$ (e.g., OTUs) is recorded in $j=1,\ldots, J$ groups (e.g., subjects).
That is, $i$ and $j$ index two different types of experimental units.
For simplicity, and without loss of generality, we will refer to column experimental units as ``groups.''
The main inference goal is to jointly identify clusters $C_{k}$, $k=1,\ldots, K$, of groups (i.e., columns) and clusters $\Psi_{j}$ of experimental units (i.e., rows) within columns -- we discuss details of $\Psi_{j}$ later.
Importantly, we want to cluster together columns with a higher probability if they share similar clusters of rows.
More precisely, the subsets $C_{k}$ define a partition as $\bigcup_{k=1}^{K} C_{k} = [J]$ with $C_{k} \cap C_{k^{\prime}}=\emptyset$ for $k \ne k^{\prime}$.
We refer to the $C_{k}$ as column clusters.
Alternatively, we represent $C_{k}$ using cluster membership indicators $\tS_{j}=k$ if $j \in C_{k}$.
We use $\tS_{j}$ when we impose an order constraint, $\tS_{j} \le \tS_{j+1}$, i.e., indexing clusters by appearance, and write $S_{j}$ without a tilde for cluster membership indicators without that constraint (throughout, we will use indicators with a tilde for cluster membership indicators in the order of appearance).
We first introduce an established partially exchangeable inference model (with also exchangeability of the groups) to argue for a variation that allows for separate exchangeability.

\paragraph*{Partially exchangeable model.}
We briefly review the partial exchangeable nested DP with common atoms (NDP-CAM) \citep{denti2023common}, which modifies the familiar nested DP (NDP) \citep{rodriguez2008nested} to allow for clustering of experimental units across different clusters of groups.
This overcomes a possible drawback of the NDP studied in \cite{camerlenghi2019latent}, which is further discussed in \cite{lijoi2023flexible, beraha2021semi, christensen2020bayes} and \cite{soriano2019mixture}.

We assume a mixture of normal distributions, but the same construction can be used for different kernel mixtures.
Let $\GEM(\alpha)$ denote a stick-breaking prior for a sequence of weights \citep{sethuraman1994constructive}.
In anticipation of the upcoming variation of the model, we start the model construction by stating the marginal model for group $j$: 
\begin{equation}
\begin{split}
    &y_{i,j} \mid S_{j}=k, G_{k} \simiid G_{k}, \, i=1,\ldots\\
    &G_{k} = \sum_{\ell=1}^{\infty} w_{k,\ell} \, \N(\mu_{\ell}, \sigma_{\ell}^{2})\\
    &\bw_{k}=(w_{k,1},\ldots) \simiid \GEM(\alpha), \, p(S_{j}=k \mid \bpi) = \pi_{k},\\
    &\bpi=(\pi_{1},\ldots) \sim \GEM(\beta), \, \theta_{\ell} = (\mu_{\ell},\sigma_{\ell}^{2}) \simiid G_{0}
\end{split}\label{eq: marg NDP-CAM}
\end{equation}
where $\btheta= (\theta_{1}, \ldots)$ are normal location and scale parameters shared across all $G_{k}$, and $\bth, \bw_{1}, \, \ldots, \,\bw_{k}, \, \bpi$ are jointly independent.
Let $K$ denote the number of unique values of the $S_{j}$.
Interpreting $S_{j}$ as cluster membership indicators, they define a partition of $[J]$ into $K$ clusters $C_{k}$ via their ties.
Both $S_{j}$ and their ordered version $\tS_{j}$ induce the same partition of the columns via their ties.
For simpler notation and to avoid empty clusters, when we state the partition as $\{C_{1},\ldots, C_{K}\}$ the running indices $k=1,\ldots, K$, are the realization of the cluster membership ordered by appearance $\tS$, i.e., $j \in C_{k}$ iff $\tS_{j}=k$ (rather than $S_{j}=k$). 
Under \eqref{eq: marg NDP-CAM}, the marginal model for $(y_{i,j};\; i=1, \ldots)$ reduces to a Dirichlet process (DP) mixture of normal models: 
\begin{align}\label{DPM}
\begin{split}
  &y_{i,j} \mid F \sim \int \N(\mu, \sigma^{2})\, \d F(\mu, \sigma^{2}) \mbox{ with }\\
  &F = \sum_{h} w_{h} \delta_{\mu_{h},\sigma_{h}^{2}} \sim \DP(\alpha,G_{0})
\end{split}
\end{align}
where $F$ can be seen as a mixture of the equally distributed unique distribution $F_{k} =\sum_{\ell} w_{k,\ell} \delta_{\mu_{\ell},\sigma_{\ell}^{2}} \sim \DP(\alpha, G_{0})$ to highlight the match with \eqref{eq: marg NDP-CAM}.
Here, $(\mu_{\ell}, \sigma_{\ell}^{2})$ are the atoms of a discrete random probability measure $F$ with a $\DP(\alpha, G_{0})$ distribution, i.e., a DP prior with total mass $\alpha$ and base measure $G_{0}$.
See, for example, \cite{muller2015bayesian}, Chapter 2, for a review of such DP mixture models.
However, the fact that in \eqref{eq: marg NDP-CAM} multiple groups can share the same $G_{k}$ introduces dependence across groups.
Additionally, note that the normal parameters $\theta_{\ell}=(\mu_{\ell},\sigma_{\ell}^{2})$ in \eqref{eq: marg NDP-CAM} are indexed by $\ell$ only, implying common atoms of the $G_{k}$ across $k$.
The common atoms across $G_{k}$ are what distinguish the NDP-CAM from the NDP.
The model construction is completed by assuming that $y_{i,j}$ in \eqref{eq: marg NDP-CAM} are conditionally independent, i.e., 
\begin{equation*}
    y_{i,j} \mid S_{j}=k, G_{k} \simind G_{k}.
\end{equation*}
The use of common atoms $(\mu_{\ell}, \sigma_{\ell}^{2})$ across $G_{k}$ allows us to have clusters of experimental units shared across clusters of groups with positive probability.
This is easiest seen by replacing the mixture of normal model in the first line of \eqref{eq: marg NDP-CAM} by a hierarchical model with latent indicators $M_{i,j}^{\dagger}$ as $p(y_{i,j} \mid M_{i,j}^{\dagger}=\ell) = \N(\mu_{\ell}, \sigma_{\ell}^{2}) \text{ and } p(M_{i,j}^{\dagger}=\ell \mid S_{j}=k) = w_{k,\ell}.$ 
 
Interpreting $M_{i,j}^{\dagger}$ as cluster membership indicators (for observations), the model defines a random partition $\Psi_{j}$ of observations for each group $j$ with clusters defined by $R_{j,\ell} = \{i:\, M_{i,j}^{\dagger}=\ell\}$, i.e., $\Psi_{j} = \{R_{j,\ell},\, \ell=1,\ldots\}$. 
In this construction, all units $j$ in the same cluster $C_{k}=\{j:\; S_{j}=k\}$ share the same {\em prior} $p(\Psi_{j})$ on the partition (rather than the $\Psi_{j}$ itself) of observations (e.g., OTUs), implied by $y_{i,j} \mid P_{j} \sim P_{j}$ with $P_{j} =G_{S_{j}}$, defining partial exchangeability as in \eqref{eq: deFPart}, with the random partition of groups obtained from the ties of $(S_{1}, \ldots, S_{J})$ introducing a dependent prior for $(P_{1},\ldots, P_{J})$ (as well as the common atoms). 
Conditional on $S_{1}, \ldots, S_{J}$ the latter is characterized by
only $K$ distinct random probability measures $G_{k}$. 

\paragraph*{A separately exchangeable prior.}
When the same experimental units, $i=1,\ldots, I$, are recorded in each group, then such a data structure calls for separate exchangeability only.
Recognizing then that the described construction with $M_{i,j}^{\dagger}$ is implementing partial exchangeability (plus exchangeability of the groups), we modify it to relax the assumptions to separate exchangeability. 
While the change is minor in notation, it has major consequences for interpretation and inference, as we will discuss later.
We replace the indicators $M_{i,j}^{\dagger}$ by $M_{i,k}$ (note the subindex $_{k}$) specific to each observation $i$ (e.g., OTU) and {\em cluster $k$ of groups}, with otherwise unchanged marginal prior 
\begin{equation*}
    p(M_{i,k}=\ell) = w_{k,\ell}
\end{equation*}
and $p(y_{i,j} \mid S_{j}=k, M_{i,k}=\ell) = \N(\mu_{\ell}, \sigma^{2}_{\ell})$.
The assumption completes the marginal model \eqref{eq: marg NDP-CAM} by introducing dependence of the $y_{i,j}$ across $j \in C_{k}$, which is parsimoniously introduced with the $M_{i,k}$ indicators.
The (per-column) marginal distribution \eqref{DPM} remains unchanged.
But now the implied random partitions $\Psi_{j}$ are shared among all $j \in C_{k}$, while before they shared just the prior law.

For later reference, we state the joint probability model of the ``occupied'' parameters $p(y_{1:I,1:J}, \bS, \bM, \bpi, \bw, \btheta)$
\begin{equation}
\begin{split}
& = \prod_{j=1}^{J} \bigg[ \pi_{S_{j}} \prod_{i=1}^{I} p(y_{i,j} \mid \mu_{M_{i,S_{j}}},\sigma^{2}_{M_{i,S_{j}}}) \bigg] \\
& \quad  \times  \prod_{k \in \{S_{j} : j \in [J]\}} \bigg[ p(\bw_{k})  \prod_{i=1}^{I} w_{k,M_{i,k}}  \bigg]\\
& \quad  \times \, p(\bpi)\, \prod_{\ell \in \{M_{i,k}: k \in \{S_{j} : j \in [J]\}, i \in [I]\} }  p(\mu_{\ell}, \sigma^{2}_{\ell})
\end{split}
\label{eq: sep rpm joint} 
\end{equation}
with $p(\bpi) = \GEM(\beta)$ and $p(\bw_{k}) = \GEM(\alpha)$ stick-breaking priors, and $p(\mu_{\ell}, \sigma^{2}_{\ell})$ chosen to be conditionally conjugate for the sampling model $p(y_{i,j} \mid \mu_{\ell},\sigma^{2}_{\ell})$.
We refer to the model as the \emph{separately exchangeable NDP-CAM}.

In summary, we have introduced separate exchangeability for $y_{i,j}$ by defining (i) a random partition $\{C_{1}, \ldots,$ $C_{K}\}$ of columns, corresponding to the cluster membership indicators $S_{j}$, and (ii) nested within column clusters $C_{k}$, a nested partition $\Psi_{j}=\{R_{j,\ell}: \, \ell=1,\ldots\}$ of rows, represented by cluster membership indicators $M_{i,k}$ with $k=S_{j}$ (i.e., $R_{j,\ell} = \{i:\, M_{i,S_{j}}=\ell\}$).
Importantly, given that two columns are clustered together, they share the same \emph{realization} of the random partition of the rows.
In contrast, a model under which $j \in C_{k}$ only shares the {\em prior} $p(\Psi_{j})$ on the nested partition reduces to the special case of partial exchangeability.
The model remains invariant under arbitrary row (OTU) label permutations.
A similar construction with nested partitions as in \eqref{eq: sep rpm joint} was also used in \cite{lee2013nonparametric} without the reference to separate exchangeability.
The construction of the nested partition is identical, but there is no notion of common atoms to allow for clusters of observations across column clusters.
In this regard, our separately exchangeable nested partition common atoms model NDP-CAM can be seen as a more flexible version of \cite{lee2013nonparametric} in the same way as \cite{denti2023common} is a more flexible version of \cite{rodriguez2008nested} for partial exchangeable nested partition models.
 
Finally, to highlight the nature of the model as being separately exchangeable with respect to rows (e.g., OTUs) and columns (e.g., groups), we exhibit the explicit Aldous-Hoover representation \eqref{th: AldHoovA}, still conditional on hyperparameters.
We show separate exchangeability conditional on $\phi=(\bpi,\bw)$ by matching variables with the quantities in \eqref{th: AldHoovA} as follows: $\eta_{j}=S_{j}$, $\xi_{i}=(M_{i,k})_{k \ge 1}$, $\theta = (\mu_{\ell}, \sigma_{\ell}^{2})_{\ell \ge 1}$, and $p(x_{i,j} \mid \theta, \xi_{i}, \eta_{j}) = \N(\mu_{\ell},\sigma^{2}_{\ell})$ with $\ell=M_{i,S_{j}}$.
Here, we used that \eqref{th: AldHoovA} allowed conditioning on additional hyperparameters, in this case, $\phi$.

\paragraph*{EPPF and pEPPF.}
We characterize the separately exchangeable random partition model \eqref{eq: sep rpm joint} directly as a composition of tractable random partition models by exhibiting the implied prior law for $\{C_{1},\ldots, C_{K}\}$ and $\Psi_{j}=\{R_{j,\ell}: \, \ell=1,\ldots\}$.
On one hand, this allows us to provide a novel understanding of model assumptions; on the other hand, it allows us to generalize the separately exchangeable NDP-CAM \eqref{eq: sep rpm joint} to different compositions of random partition models, making use of recent advances in the BNP literature.
Let then $\bN=(N_{1},\ldots, N_{K})$ denote the cluster sizes $N_{k}=|C_{k}|$ of the column clusters.
The nature of \eqref{eq: marg NDP-CAM} as a DP mixture model implies that the column partition induced by $\bS$ is characterized by the exchangeable partition probability function (EPPF) related to the Chinese restaurant process, i.e., the EPPF for the partition generated by the ties under i.i.d.\ sampling from a DP random measure:
\begin{align}\label{eq: EPPF Dir}
\begin{split}
    &\Pr\{C_{1},\ldots,C_{K}\} = \mbox{EPPF}_{K}^{(J)}(N_{1},\ldots,N_{K})\\
    &\quad =\frac{\beta^{K}\Gamma(\beta)}{\Gamma(\beta+J)} \prod_{k=1}^{K} (N_{k}-1)!
\end{split}
\end{align}
The EPPF characterizes the distribution of an exchangeable partition \citep{pitman1996some}, with $\mbox{EPPF}_{K}^{(J)}(N_{1},\ldots, N_{K})$ being the probability of observing a particular (unordered) partition of $J$ objects into $K$ subsets of cardinalities $\{N_{1},\ldots, N_{K}\}$.

Given the realization of the column random partition, $\{ C_{1}, \ldots, C_{K} \}$, the random partition of the rows is equal in all columns that are clustered together.
That is, $\Psi_{j} = \Psi_{k}^{\ast}$ for each $j \in C_{k}$ and $k \in [K]$. 
Here, $\Psi_{k}^{\ast}$ denotes the unique random partitions of the rows (i.e., $p(\Psi_{k^{\prime}}^{\ast} \ne \Psi_{k}^{\ast})>0$ iff $k \ne k^{\prime}$).

Recall that the cluster membership indicators $M_{i,k}$ in the separately exchangeable NDP-CAM \eqref{eq: sep rpm joint} can include ties across column clusters $k$.
As a consequence, row clusters can be shared across different column clusters $k$.
Let $R \le I\cdot K$ denote the number of unique values among the $M_{i,k}$ indicators, and let $\tM_{i,k} \in [R]$ denote the indicators after re-indexing the row clusters by appearance such that $\tM_{i,k} \le \tM_{i+1,k}$ and $\tM_{I,k} \le \tM_{1,k+1}$ (i.e., ordering by appearance column cluster by column cluster).
Note that $\bM$ and $\tbM$ induce the same partition structure of the rows, similarly to $\bS$ and $\tbS$ for the column clusters.
Let then $\bn_{k}=(n_{k,1},\ldots,n_{k, R})$ denote the frequencies of the unique values in $\{\tM_{1,k},\ldots,\tM_{I,k}\}$.
For example, $n_{k,r}$ indicates the number of elements in the $k$th column cluster that coincide with the $r$-th distinct value in the order of arrival. 
Clearly, $n_{k,r} \ge 0$ and $\sum_{k=1}^{K} n_{k,r} \ge 1$.
One may well have $n_{k,r}=0$, which implies that the $r$th distinct value is not recorded in the $k$th cluster of columns, though by virtue of $\sum_{k=1}^{K} n_{k,r} \ge 1$ it must be recorded at least in one of the clusters of columns.
The $r$th distinct value is shared by any two groups $k$ and $k^{\prime}$ if $n_{k,r} \ge 1$ and $n_{k^{\prime},r} \ge 1$.

We conclude the characterization of the separately exchangeable random partition by explicitly stating the probability function for the nested partially exchangeable partition of the rows.
Here, the groups for conditional partial exchangeability are the clusters of columns.
We give the corresponding partial exchangeable random partition function (pEPPF) \citep[see e.g.,][]{franzolini2025multivariate}, which is the natural generalization of the concept of EPPF for the exchangeable case.
Given the clustering of the columns from \eqref{eq: EPPF Dir}, the probability law of the random partition structure of the rows (i.e., $\Psis_{1}, \ldots, \Psis_{K}$) is characterized by the pEPPF defined as follows.
Let $(\tbw_{1},\ldots,\tbw_{K})$ denote the unique values of $ (\bw_{S_{1}},\ldots,\bw_{S_{J}})$ in order of arrival given $\bS$.
Then, given $\bS$,
\begin{equation}\label{eq: pEPPF}
    \mathrm{pEPPF}_{R}^{(N)}(\bm{n}_{1},\ldots,\bm{n}_{K}) = \mathbb{E} \bigg[ \sum_{h_{1} \ne \ldots \ne h_{R}} \prod_{k=1}^{K} \prod_{r=1}^{R} \tw_{k,h_{r}}^{n_{k,r}} \bigg].
\end{equation}
with $N=I \cdot K$ and the constraint $\sum_{r=1}^{R} n_{k,r} = I$, for each $k=1,\ldots,K$.
The expected value in \eqref{eq: pEPPF} is computed with respect to the joint law of the vector of the random probabilities $(\tbw_{1},\ldots,\tbw_{K})$.
Note that the pEPPF in \eqref{eq: pEPPF} is the one induced by the ties between parameters arising from the dependent discrete non-parametric probabilities $(\tF_{1},\ldots, \tF_{K})$ defined by $\tbw_{k}$, i.e., that are defined as the unique random probabilities in order of arrival among $(F_{S_{1}}, \ldots, F_{S_{J}})$.
That is, $\tF_{k}=\sum_{\ell} \tilde{w}_{k,\ell} \delta_{\mu_{\ell},\sigma_{\ell}^{2}}$ (note that the pEPPF does not depend on $\mu_{\ell},\sigma_{\ell}^{2}$).
In the special case of a single column cluster (i.e., $K=1$), the standard EPPF induced by the DP (i.e., \eqref{eq: EPPF Dir}) is recovered.

Together, by first sampling the random column partition from \eqref{eq: EPPF Dir} and then the nested row partitions from \eqref{eq: pEPPF}, we define the separately exchangeable partition implied by \eqref{eq: sep rpm joint}.
The model construction can then be concluded by i.i.d.\ sampling of the unique atoms $(\tht_{m} = (\mut_{m}, \sigst_{m}))_{m=1}^{L}$ from $G_{0}$ (the $\tht_{m}$ are a reordering of a subset of the $\theta_{\ell}=(\mu_{\ell},\sigma_{\ell}^{2})$ in \eqref{eq: sep rpm joint}), and finally sample $y_{i,j} \sim \N(\tmu_{m},\sigst_{m})$, $i \in R_{jm}$, independently across $i$ and $j$.

This characterization of the model as a random partition model prepares for the introduction of marginal posterior simulation schemes and upcoming generalizations, while still maintaining analytical and computational tractability.

\begin{table*}
    \caption{$p(\tilde{S}_{j}=k \mid \ldots)$ in the gCRP \eqref{eq: Gibbs urn}, and associated weights $(\pi_{k})_{k=1}^{K^{+}}$ that generalize the ones in \eqref{eq: sep rpm joint}.
    See the text for details on the four models described below.}
  \begin{center}
    {\renewcommand{\arraystretch}{1.5}
    \scriptsize
      \begin{tabular}{c|l:l|l|l}
        & \multicolumn{2}{l|}{\hspace{.5cm}
          $p(\tilde{S}_{j}=k \mid \tilde{\bS}^{-j})\propto$} 
        & $P(\pi_{1}, \ldots, \pi_{K^{+}} \mid K^{+}) $
        & $p(K^{+}=k)$\\
        { Gibbs-type prior}
        & $k \in \tilde{\bS}^{-j}$
        & $k = K^{-j} +1$ & \\
        \hline
        $K^{+}-$dim symm Dir %
        & $N_{k}^{-j} + \rho $
        & $\rho (K^{+}-K^{-j})\;^{(a)}$ 
        & $ \Dir(\rho,\ldots,\rho)$ 
        & $K^{+} \in \mathbb{N}$ \\
        MFM: Gnedin (discount$=-1)$%
        & $(N_{k}^{-j}+1)\; \times$ 
        & $(K^{-j})^{2}-K^{-j} \gamma$
                                                        & $\Dir(1,\ldots,1)$ 
                      & $\frac{\gamma (1-\gamma)_{k-1}}{k!}$\\
         \citep[with $\zeta=0$ in][]{gnedin2010species}
        & $~~ (J^{-j}-K^{-j}+\gamma)$ 
        & 
                                                        & 
                      & \\
        DP
        & $ N_{k}^{-j}$ 
        & $\beta$ 
                                                        & $\GEM(\beta)\; ^{(b)}$
                      & $K^{+}=\infty$ \\
        PYP (with $\sigma>0$)
        & $N_{k}^{-j}-\sigma$
        & $ \beta +K^{-j} \sigma $ 
                                                        & $\GEM(\beta,\sigma)\; ^{(b)}$ & $K^{+}=\infty$
      \end{tabular} }
  \end{center}
        $^{(a)}$ subject to $K^{-j}<K^{+}$.
        $^{(b)}$ $\GEM$ stands for the distribution of probability weights after Griffiths, Engen, and McCloskey \citep{ewens1990population}, using the 1-parameter version defined there and the related 2-parameter extension.
	\label{tab: PPF}
\end{table*}
\subsection{Models beyond the DP prior}
A natural first generalization is to substitute the EPPF in \eqref{eq: EPPF Dir} with a different prior for the column partition $\{C_{1},
\ldots, C_{K}\}$ to allow for increased flexibility in representing prior beliefs. 
The law of the random partition induced by the DP, related to the so-called Chinese restaurant process (CRP), is controlled by a single parameter (see Table \ref{tab: PPF} for details about the CRP).
This leaves DP mixture models too restrictive for some applications, leading to several alternative models being introduced in the literature.
This includes the symmetric finite Dirichlet prior \citep{green2001modelling}, the Pitman-Yor process (PYP) \citep{pitman1997two}, the normalized inverse Gaussian (NIG) \citep{lijoi2005hierarchical}, the normalized generalized gamma process (NGGP) \citep{lijoi2007controlling}, mixture of finite mixtures (MFM) \citep{nobile1994bayesian, richardson1997bayesian, nobile2007bayesian, miller2018mixture} and the mixture of DP (MDP) models \citep{antoniak1974mixtures, ascolani2023clustering}.
A common framework for all these is the family of Gibbs-type priors \citep{gnedin2006exchangeable} that can be seen as a natural, flexible generalization of the DP \citep{deblasi2015gibbs}.
The Gibbs-type prior EPPF generalizes \eqref{eq: EPPF Dir} to 
\begin{equation}\label{eq: EPPF Gibbs}
    \mbox{EPPF}_{K}^{(J)}(N_{1},\ldots,N_{K} \mid \beta, \sigma) = W_{J,K} \prod_{k=1}^{K} (1-\sigma)_{N_{k}-1},
\end{equation}
where $(x)_{J}=x(x+1)\ldots(x+J-1)$ represents the ascending factorial, $\sigma<1$ is a discount parameter and the set of non-negative weights $\{W_{J,k}: 1 \le k \le J \}$ satisfies the recursive equation $W_{J,k} = (J - \sigma k) W_{J+1,k} + W_{J+1,k+1}$.

Sampling from the EPPF for the Gibbs-type prior \eqref{eq: EPPF Gibbs}
is implemented via sampling of the cluster membership indicators $\tilde{\bS}=(\tilde{S}_{1},\ldots, \tilde{S}_{J})$ recursively from the corresponding generalized CRP (gCRP).
We write $\tbS \sim \text{gCRP}$ defined as
\begin{equation} \label{eq: Gibbs urn}
    p(\tilde{S}_{j}=k \mid \tilde{\bS}^{-j}) =
    \begin{cases}
	\frac{W_{J,K^{-j}}}{W_{J-1,K^{-j}}} (N_{k}^{-j}-\sigma) & k \in [K^{-j}]
	\vspace*{0.2cm}\\
	\frac{W_{J,K^{-j}+1}}{W_{J-1,K^{-j}}} & k = K^{-j} +1.
    \end{cases}
\end{equation}
Throughout $\mathbf{x}^{-j}$ identifies a quantity after removing the element $j$ from $\mathbf{x}$.
Recall that $\tilde{\bS}$ is just a relabeling (in order of arrival) of $\bS$ such that $\tilde{S}_{1}=1$ and $\tilde{S}_{1} \le \tilde{S}_{2} \le \ldots \tilde{S}_{J}$.
For several Gibbs-type priors \eqref{eq: Gibbs urn} reduces to very simple ratios (see Table \ref{tab: PPF}) that allow the use of marginal algorithms \citep[see e.g.,][]{neal2000markov, deblasi2015gibbs, miller2018mixture} and analytical results similar to the ones of the DP.
Importantly, Table \ref{tab: PPF} also recalls the distributions of the weights $\bpi$ that allow us to adapt the conditional algorithm described in Section S.3 of the supplementary materials, or extend other conditional algorithms proposed in the literature \citep{escobar1994estimating, blei2017variational, lijoi2023flexible, giordano2023evaluating}.

Besides generalizing the EPPF in \eqref{eq: EPPF Dir}, another natural generalization of the separately exchangeable NDP-CAM \eqref{eq: sep rpm joint} is to substitute the pEPPF \eqref{eq: pEPPF}.
Substituting a different pEPPF in \eqref{eq: pEPPF} is equivalent to replacing the independent $\mbox{GEM}(\alpha)$ in \eqref{eq: sep rpm joint} by a different prior for $(\tbw_{1},\ldots,\tbw_{K})$, or replacing the CAM DDP by another dependent prior for $\tF_{1}, \ldots, \tF_{K}$.

\paragraph*{Separately exchangeable NDP without common atoms.}
A simple example is to define the prior for $\tF_{1},\ldots, \tF_{K}$ as independent DP without common atoms, or, equivalently, the pEPPF such that marginally in each $k$ we have the EPPF of a DP \eqref{eq: EPPF Dir}.
In this way, we obtain the model in \cite{lee2013nonparametric} that is the separately exchangeable counterpart of the partially exchangeable NDP \citep{rodriguez2008nested}.

\paragraph*{Separately exchangeable NDP-CAM.}
We note that recently several nested partial exchangeable models were introduced to have more flexible nested random partition models \citep[see e.g.,][]{camerlenghi2017bayesian, denti2023common, lijoi2023flexible, beraha2021semi}.
We recall that our proposal is the separately exchangeable counterpart of \cite{denti2023common}, and now briefly comment on how to define flexible counterparts of other recently proposed BNP/random partition models, by replacing the prior for $(\tF_{1},\ldots,\tF_{K})$ by another family of dependent processes, or, equivalently, replacing their pEPPF.
\paragraph*{Separately exchangeable hidden hierarchical DP (HHDP).}
An interesting example is to choose a hierarchical Dirichlet process (HDP) \citep{teh2006hierarchical} prior for $(\tF_{1},\ldots,\tF_{K})$, or equivalently, use the pEPPF of the HDP \citep{camerlenghi2019distribution} related to the well-known and tractable Chinese restaurant franchise, allowing us to define the separately exchangeable counterparts of the HHDP \citep{lijoi2023flexible}.

\paragraph*{Separately exchangeable latent nested process (LNP)}
Setting $(\tF_{1},\ldots,\tF_{K})$ as dependent processes arising from additive structure between distributions with a common component and one specific of the distribution $k \in [K]$ as done in \cite{muller2004method} or \cite{lijoi2014bayesian} creates the separately exchangeable counterpart of the LNP \citep{camerlenghi2019latent}.

\paragraph*{Separately exchangeable semi-HDP}
Finally, choosing a prior for $(\tF_{1},\ldots,\tF_{K})$ by combining in an additive way the HDP and a non-atomic measure as done in \cite{beraha2021semi} or, equivalently, specifying the pEPPF in \eqref{eq: pEPPF} as in \cite{beraha2021semi} allows specifying the separately exchangeable counterpart of the semi-HDP. 

Similar ideas can be applied to other tractable pEPPF or dependent processes available in the literature or even to new tractable ones. 

 
\section{Separate Exchangeability in Nonparametric\\
Regression}
\label{sec: sep ANOVA}
We set up a separately exchangeable prior for BNP regression as an implementation of the popular DDP.
The desired separate exchangeability is achieved by means of introducing the symmetric structure with respect to two types of experimental units in the prior for the atoms in a DP random measure.
For example, the separate exchangeability assumption could be on protein expression profiles (over time) for proteins $i=1, \ldots, I$ and patients $j=1, \ldots, J$.
The actual construction is simple and tractable.
In short, we achieve separate exchangeability by replacing atoms under the DDP model with linear predictors including row and column-specific terms.
 
Recall that the DDP is a predictor-dependent extension of DP priors that defines a family of random probability measures $\F=\{F_{x}: \, x \in X\}$, where the random distributions $F_{x}$ are indexed by covariates $x$ and each $F_{x}$ marginally follows a DP prior as in \eqref{DPM}. 
The desired dependence of $F_{x}$ across $x$ is achieved by modeling the atoms and/or the weights of the DP random measure $F_{x}$ as functions of covariates \citep{maceachern1999dependent, maceachern2000dependent, quintana2022dependent}.
The most commonly used variation of DDP models assumes common weights and varying, predictor-dependent atoms.
If desired, a continuous sampling model is constructed using an additional convolution with, for example, a Gaussian kernel. 
The model for $\{y(x): \, x \in X\}$ becomes
\begin{equation*}
    y_{i} \mid \bx_{i}=\bx, \mathcal{F} \simind F_{\bx} \mbox{ and }
    F_{\bx} = \sum_{h} \pi_{h} \N(\beta_{h}(\bx), \sigma^{2}),
\end{equation*}
defining a prior for $\FF=\{F_{\bx}: \, \bx \in X\}$, with, for example, a Gaussian process prior for $\beta_{h}(X) = (\beta_{h}(\bx), \bx \in X)$.
We conclude the model specification with a prior for the common hyperparameter, $\sigma^{2} \sim p_{\sigma}$.

A simple instance of the DDP is the ANOVA DDP \citep{deiorio2004anova} that arises when the atoms of $F_{\bx}$ are specified as in an ANOVA model.
The model can be written as a DP mixture of linear models, that is, as a mixture with respect to some (or all) linear model parameters, and a DP prior for the mixing measure.
The ANOVA DDP defines a simple nonparametric regression.
Let $\bd=\bd(\bx)$ denote a known design vector.
The ANOVA DDP model assumes 
\begin{equation}
  y_{i} \mid \bx_{i}=\bx, \mathcal{F} \sim F_{\bx} \mbox{ and }
  F_{\bx} = \sum_{h} \pi_{h} \N(\beta_{h}^{\top} \bd, \sigma^{2}).
\label{DDP}
\end{equation}
Alternatively, many implementations of BNP regression are based on basis expansions, using, for example, a spline-based representation for a regression mean function plus a parametric residual.
But in the following construction, we build on the DDP model as a widely used BNP regression framework.

Model \eqref{DDP} can be naturally adapted to a separately exchangeable structure.
Let $\by_{i,j}=y_{i,j}(X)=(y_{i,j}(x): x \in X)$ denote potential outcomes for experimental units $i=1, \ldots, I$ and $j=1,\ldots,J$ as a function of covariates $x \in X$.
Posterior inference will eventually condition on the observed values $y_{i,j}(x)$ for a subset of $x$ values.
But for the discussion of separate exchangeability, we consider a probability model for the entire profile $\by_{i,j}$.
Assume then, for a moment, that $X$ is finite.
For example, in a later illustration $x=(t,z)$ includes age $t$ and an indicator $z$ for case versus control, defining $\by_{i,j}$ to be the (potential) protein expression profile for protein $i$ in patient $j$ across age $t$ and condition $z$.
Let $\bd_{\beta}=\bd_{\beta}(\bx), \bd_{\xi}=\bd_{\xi}(\bx)$ and $\bd_{\eta}=\bd_{\eta}(\bx)$ denote fixed design vectors.
We modify the ANOVA DDP \eqref{DDP} to $y_{i,j}(\bx) \mid \mathcal{F} \sim F_{\bx}\ij$ with $F_{\bx}\ij = \sum_{h} \pi_{h} \N(\bd_{\beta}^{\top}\beta_{h} + \bd_{\xi}^{\top} \xi_{i} + \bd_{\eta}^{\top} \eta_{j}, \sigma^{2})$.
For a finite covariate space, we can collect the design vectors across $x \in X$ in design matrices $\bD_{\beta}, \bD_{\xi}$ and $\bD_{\eta}$, with one row for each unique $x$, and rewrite the model as
\begin{align}\label{eq: ANOVA lin}
\begin{split}
  &\by_{i,j} \mid \Pij \sim \Pij \mbox{ with }\\
  &\Pij = \sum_{h} \pi_{h}
  \N(\bD_{\beta}\beta_{h} + \bD_{\xi} \xi_{i} + \bD_{\eta}\eta_{j},
  \sigma^{2}\, I).
\end{split}
\end{align}
The choice of the design matrices depends on the desired inference and on the available data.
Importantly, the model entails separate exchangeability on $\by$ via \eqref{th: AldHoovA} if the model is completed with a prior 
\begin{equation}\label{eq: prior ANOVA lin}
  \btheta =((\pi_{h}, \beta_{h})_{h}, \sigma^{2}) \sim p_{\theta} \perp
  \xi_{i} \simiid p_{\xi} \perp
  \eta_{j} \simiid p_{\eta}.
\end{equation}
Under the DDP prior $p_{\theta}$, $\bpi=(\pi_{1}, \pi_{2}, \ldots) \sim \GEM(\alpha)$ and $\beta_{h} \simiid P_{0}$ for some base distribution in \eqref{eq: prior ANOVA lin}.
And $p_{\xi}$ and $p_{\eta}$ can be any probability models, possibly including additional hyperparameters or with a nonparametric prior themselves.

We assumed a finite covariate space $X$ for easier notation, but the same model structure could be assumed for a more general covariate space.
For example, for $X = \R^{p}$ the multivariate normal in \eqref{eq: ANOVA lin} would be replaced by a Gaussian process with an independent covariance kernel.

Another generalization is motivated by the implied DP prior for $P=\sum_{h} \pi_{h} \delta_{\beta_{h}}$. 
Consider a hierarchical model representation of \eqref{eq: prior ANOVA lin} with latent $\tbeta_{i,j}$:
\begin{align*}
    &\by_{i,j}\mid \tilde{\beta}_{i,j}, \tilde{\eta}_{j}, \tilde{\xi}_{i},
    \tilde{\Sigma}_{i,j} \simind
    \N(\bD_{\beta}\tilde{\beta}_{i,j} +\bD_{\xi} \tilde{\xi}_{i}
    + \bD_{\eta} \tilde{\eta}_{j},
    \, \tilde{\Sigma}_{i,j})\\
    & \betat_{i,j} \mid P \simiid P, \,\
    P = \sum_{h} \pi_{h} \delta_{\beta_{h}} \sim \DP(\alpha, P_{0}),
\end{align*}
where $\tilde{\xi}_{i} \simiid p_{\xi} \perp \tilde{\eta}_{j} \simiid p_{\eta}$ as before and independent from $\tilde{\Sigma}_{i,j}$, where we set $\tilde{\Sigma}_{i,j} \coloneqq \sigma^{2} I$, with $\sigma^{2} \sim p_{\sigma^{2}}$.
Stating the model as a DP mixture suggests possible generalizations similar to the nested partition models discussed earlier, by substituting the DP prior for $P$ with another more flexible Gibbs-type process, such as the PYP.
Importantly, this can be done while preserving computational tractability thanks to the simple gCRP described in Table \ref{tab: PPF}.

Moreover, to have a flexible separately exchangeable latent clustering (across $i$ and $j$), we can use additional discrete Gibbs processes as hyper-prior for $p_{\xi}$ and $p_{\eta}$, e.g.
\begin{equation}\label{PYP-DDP} 
    p_{\xi} \sim \text{PYP}(\alpha_{\xi}, \sigma_{\xi}, P_{0,\xi}), \,\
    p_{\eta} \sim \text{PYP}(\alpha_{\eta}, \sigma_{\eta}, P_{0,\eta}).
\end{equation} 
(or also as an alternative to the DP prior for $P$).
We shall use \eqref{PYP-DDP} in the examples in \ref{sec: illu anova data}.
See also S.4 of the supplementary materials for the characterization of such models in terms of compositions of simple urn schemes that allow us to derive a tractable marginal Gibbs sampler.

In summary, we introduced a separately exchangeable structure in a DDP regression model by simply adding additive effects in the atoms of DDP random measures.
The construction becomes especially simple in the case of the ANOVA DDP, for which we stated it in \eqref{eq: ANOVA lin}, but many alternative constructions to the additive structure that still fit into the representation \eqref{th: AldHoovA} are possible.
For example, instead of an additive structure, one could use matrix factorizations with row- and column-specific factors \citep[e.g.,][]{guhaniyogi2017bayesian}.

\section{Examples}\label{sec: illu}
We illustrate the use of the proposed separately exchangeable models for nested partitions and BNP regression in two examples, focusing on setting up the model structure, rather than extensive data analysis.

\subsection{Microbiome data}
We analyze publicly available microbiome data from a diet swap study \citep{okeefe2015fat} which includes microbiome data for $J=38$ subjects and $I=119$ OTUs (operational taxonomic units).
The data reports the frequencies $y_{i,j} \in \mathbb{R}$ of OTU $i$ for subject $j$ (after standard pre-processing that entails discarding underrepresented OTUs and the normalization and log transformation of the frequencies).
The dataset includes subjects from the US and rural Africa.
Our main goal is to investigate the different patterns of microbial diversity (OTU counts) across subjects and how such patterns vary across subgroups of subjects.
Some summaries of the data appear in Section S.1 of the supplementary materials.

The same data was analyzed by \cite{denti2023common}.
Focusing on inference for microbial diversity, they restricted attention to inference about identifying clusters of subjects with similar distributions of OTU frequencies, considering OTU counts $y_{i,j}$ as partially exchangeable.
While this focus is in keeping with the tradition in the relevant literature, it ignores the shared identity of the OTUs $i$ across subjects $j$.
Using model \eqref{eq: sep rpm joint}, we implement alternative inference on clustering subjects with respect to microbiome profiles while accounting for the separately exchangeable structure of the data.
That is, we set up inference that respects OTU identity by modeling the data as separately exchangeable with respect to subject indices $j$ and OTU indices $i$.

Posterior simulation is implemented as a partially collapsed Gibbs sampling algorithm (see Algorithm 1 in the supplementary materials for details).
Posterior Monte Carlo simulation is followed by a summary of the random partition that minimizes the expected posterior value of the variation of information loss (VI) \citep{meila2007comparing} as suggested by \cite{wade2018bayesian} and implemented in \texttt{salso} \citep{dahl2022salso}.
Alternative loss functions can be used as needed for different applications, such as the Binder loss (BI) \citep{binder1978bayesian} or an entropy-regularized version of VI or BI losses \citep{franzolini2024entropy}.
Finding a point estimate for multilevel partitions, such as nested partitions in partially and separately exchangeable models, can be challenging.
For more details and discussion on the topic, see Section S.5 of the supplementary materials.

\begin{figure}
    \centering
    \vspace{-3mm}
    \includegraphics[width=0.45\textwidth]{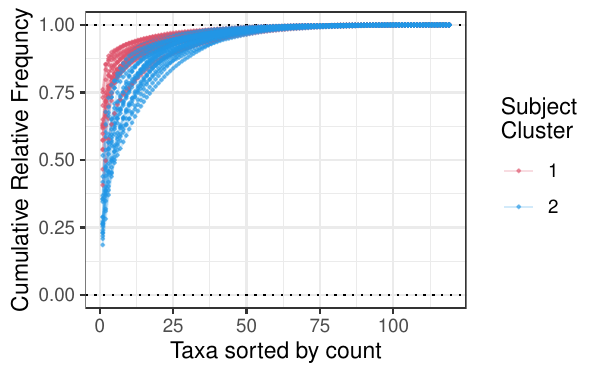}
    \caption{Cumulative relative frequencies of OTUs in the subject clusters $C_{k}$, $k=1,2$, of subjects.}
\label{fig: Sj-mb}
\end{figure}

We estimate three clusters of subjects, $C_{1}, C_{2}, C_{3}$.
The latter, $C_{3}$, is a single subject that is an outlier with very different OTU values. 
We show the other two subject clusters in Figure~\ref{fig: Sj-mb} by plotting the cumulative frequencies of OTUs.
We sort all OTUs by overall abundance across subjects.
For each cluster of subjects, we collect all subjects $j$ and plot their cumulative (observed) frequencies $\mbox{CRF}_{j}(i)\coloneqq \sum_{l=1}^{i} z_{(l),j}/ \sum_{l=1}^{I} z_{(l),j}$ where $z_{(l),j}$ denotes the $\ell-$th highest frequency among the observed OTUs in subject $j$ (recall that the $y_{i,j}$ are obtained by pre-processing $z_{i,j}$).
The results align with the distributional clusters of subjects reported by \cite{denti2023common}, who assumed a partially exchangeable NDP-CAM. 
The two subject clusters divide the subjects quite well into the US and rural Africa, as expected. 
Note, as in \cite{denti2023common}, we held out such information in the data.

Contrary to inference under a partially exchangeable framework, we can report meaningful inference about the clustering of OTUs within subject clusters, in a way that respects OTU identity. 
Similarly, subject clusters are defined in a way that accounts for the available information on OTU identity and are not based on only the marginal distributions as discussed before. 

\begin{figure*}
  \centering
  \includegraphics[width=0.7\textwidth]{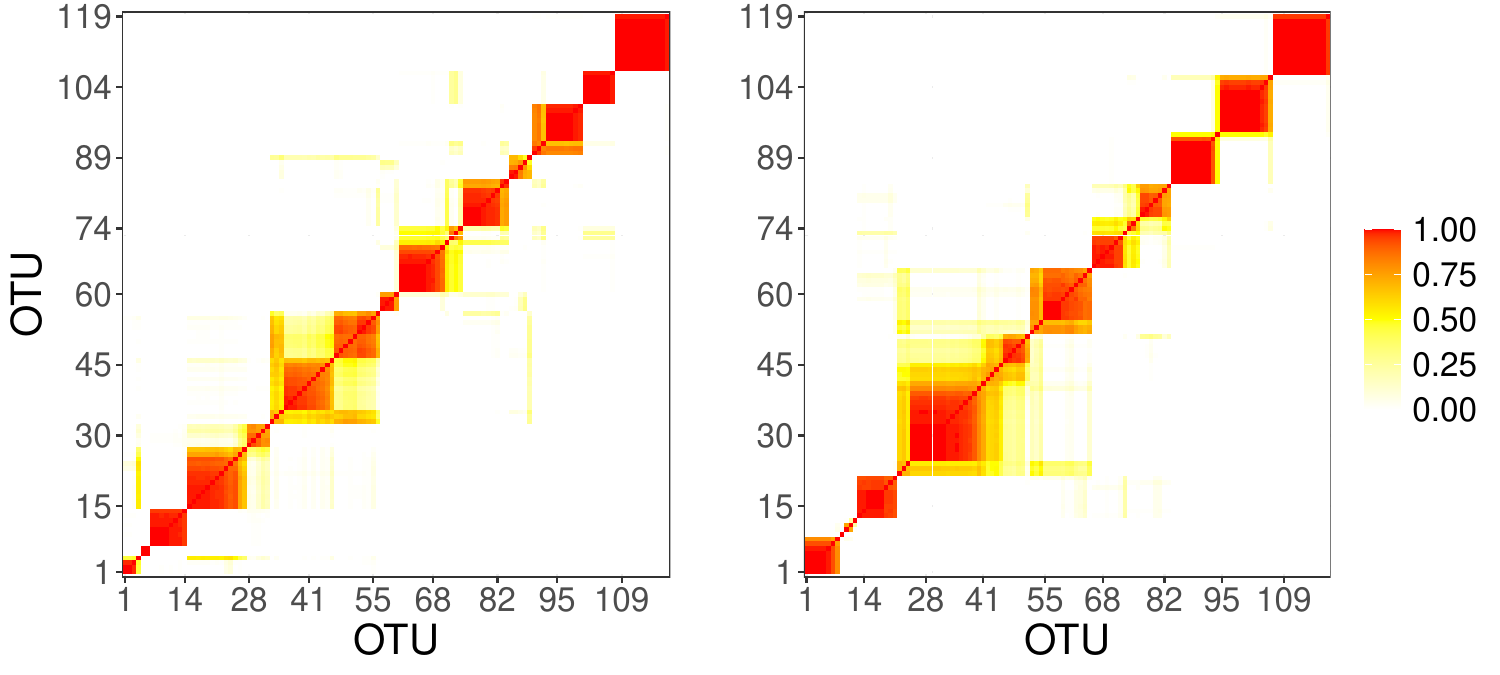}
  \caption{OTU co-clustering probability under clusters of subjects $k=1$ (left panel) and $k=2$ (right panel).
    In each block, OTUs are ordered by their cluster assignment.}
  \label{fig: otu-cocluster}
\end{figure*}

Figure \ref{fig: otu-cocluster} summarizes the nested partition of OTUs, nested within the three subject clusters.
The two panels correspond to subject clusters $k=1,2$.
For each subject cluster, the figure shows the estimated co-clustering probabilities of OTUs, i.e., $p^{k}_{i,i^{\prime}}=p(M_{i,k}=M_{i^{\prime},k} \mid \by, \bS)$ for each pair $(i,i^{\prime})$ of OTUs.
As usual for heatmaps, OTUs are sorted for better display, to highlight the clusters.
Figure~S.3 in the supplementary materials shows the same nested partitions, but now by showing the data $y_{i,j}$ arranged by subject clusters.

\subsection{Protein expression profiles}
\label{sec: illu anova data}
\begin{figure*}
    \centering
    \includegraphics[height=0.26\linewidth]{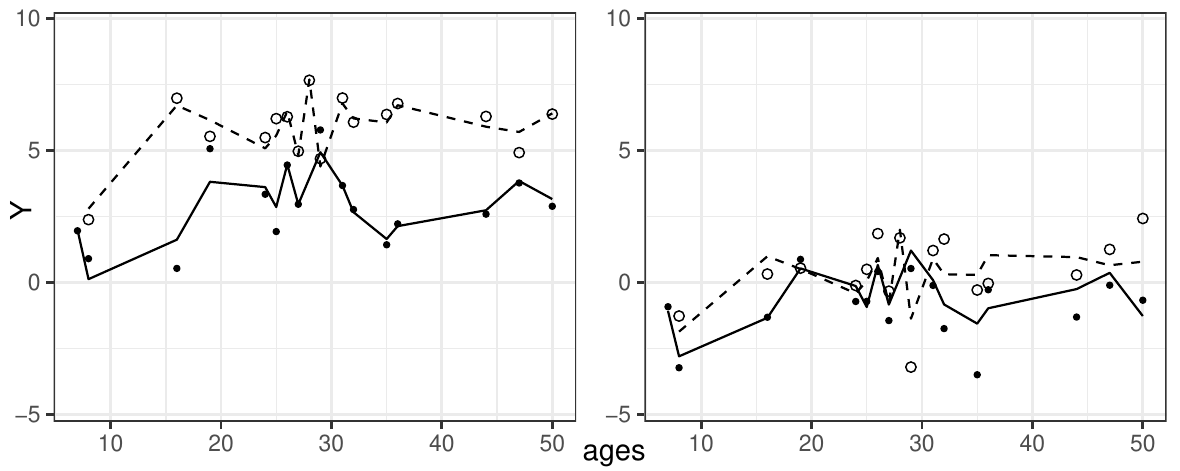}
    \includegraphics[height=0.26\linewidth]{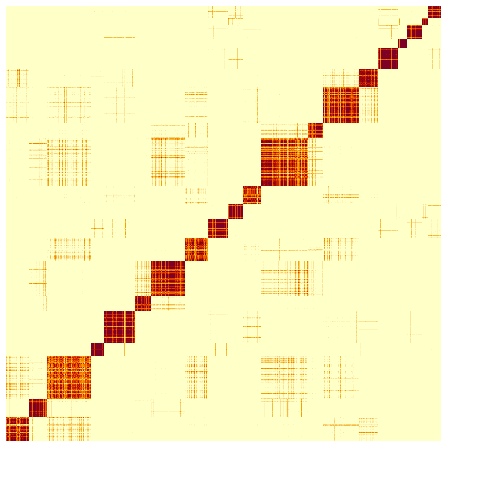}
    \caption{Left Panel: Plot of estimated protein profiles under ataxia and control (solid and dashed lines, respectively) for two randomly selected proteins.
    Points represent observed data for ataxia and control (solid and empty dots, respectively).
    Right Panel: Co-clustering probabilities for proteins.
    \label{fig: prot profile}}
\end{figure*}

In a second example, we analyze protein activation data in a study of ataxia, a neurodegenerative disease.
The same data is studied in \cite{ryu2020proteome}.
The data measures the abundance of $4350$ potentially disease-related proteins among two groups of subjects, one control group and one patient group.
Each group includes $16$ subjects between $5$ and $50$ years of age.
Measurements for two (arbitrarily selected) proteins are shown in the left panel of Figure \ref{fig: prot profile}.
Dropping proteins with fewer than $10$ recorded values leaves $3990$ proteins in the data set.
The measurements are recorded in a data matrix with $y_{i,j} \in \mathbb{R}$ denoting the abundance of protein $i$ in subject $j$ collected across the $I=3990$ proteins and $J=32$ subjects.
The measurements include some missing values, which are assumed to be missing at random.
The data includes two subject-specific covariates: an indicator $z_{j} \in \{0,1\}$ for ataxia ($1$) versus control ($0$), and age (in years) $t_{j} \in A_{T} \equiv \{7, 8, 16, 19, 24, 25, 26, 27, 28, 29, 31, 32, 35, 36, 44, 47, 50\}$ with $T=17$ unique age values under all conditions; $t_{j}=7$ and $t_{j}=28$ are only recorded for controls and cases, respectively.
For later reference, let $A_{T_{1}}$ denote the $16$ unique values for cases and $A_{T_{0}}$ the same for controls.
In summary, the design of the study was to record all proteins for all patients, but only one patient for each age and condition.
The primary aim is to identify proteins related to the disease, defined as proteins with a large difference between patients and controls in changes of protein expression over age.
With this goal in mind, we set up a nonparametric regression for protein expression versus age, under each of the two conditions.
The study design calls for a prior model with separate exchangeability with respect to patient and protein indices.
We implement an instance of the ANOVA DDP model \eqref{eq: ANOVA lin}:
\begin{align}\label{py-prot}
\begin{split}
  &y_{i,j}(X) \mid \xi_{i}, \eta_{j}, \sigma^{2} \sim \N(\eta_{j} + \bD_{\xi} \xi_{i}, \, \sigma^{2} I),\\
  &\sigma^{2} \sim \text{Inv-Ga}(a_{\sigma}, b_{\sigma})
\end{split}
\end{align}
with PYP prior \eqref{PYP-DDP}, using base measures $P_{0,\xi}= \N_{12}(\mu_{\xi}, \Sigma_{\xi})$ and $P_{0,\eta} = \N(\mu_{\eta}, \sigma_{\eta}^{2})$.
Recognizing the limited information in the data, with only one design point $(t_{j},x_{j})$ per subject, we use $D_{\beta}=\emptyset$ and $D_{\eta}=1$ in \eqref{py-prot}.
We use the design matrix $\bD_{\xi}$ to implement a cubic B-spline with two internal nodes to represent protein expression profiles over $A_{T_{0}}$ and $A_{T_{1}}$ for controls and cases, respectively.
That is, the rows of $\bD_{\xi}$ are cubic B-spline basis functions (with $2$ internal knots) evaluated at ages $t_{j} \in A_{T_{0}}$ for
rows corresponding to controls ($z=0$) and $t_{j} \in A_{T_{1}}$ for rows corresponding to cases.
More specifically, let $B_{i}(t)$ denote the $i$th cubic B-spline basis function evaluated at age $t$.
For $x=(z=0,t)$, i.e., for controls, the corresponding row of $\bD_{\xi}$ is the design vector $\bd(0,t)=(B_{1}(t), \ldots, B_{6}(t), 0, \ldots, 0)$, and for cases $\bd(1,t)=(B_{1}(t), \ldots, B_{6}(t), B_{1}(t), \ldots, B_{6}(t))$. 
This implies an interpretation of $\xi_{i} \in \R^{12}$ with the first
6 coefficients indexing the protein expression profile for controls,
and the last 6 coefficients indexing an offset for cases. 
Discussion on the choice of the hyperparameter setting in the analysis and the MCMC sampler to perform inference is reported in Section S.4 of the supplementary materials.

Recall that the main inference goal is to identify proteins with large differences (across cases and controls) in protein expression's overall slope (with respect to age).
Let $t_{T}=50$ and $t_{1}=8$ denote the age range (for cases).
Let $a_{7:12}$ denote the last 6 elements of a vector $a \in \R^{12}$.
We define $\gamma_{i} = [\bd(1,t_{T})-\bd(1,t_{1}) - (\bd(0,t_{T})-\bd(0,t_{1}))] \xi_{i} = [\bd(1,t_{T})_{7:12} - \bd(1,t_{1})_{7:12}] (\xi_{i})_{7:12}$ as the difference of slopes (ignoring division by $t_{T}-t_{1}$) for protein $i$.
We cast the problem of identifying interesting proteins as a problem of ranking proteins by $|\gamma_{i}|$, and more specifically, one of estimating a certain quantile, to report the most promising $100(1-c)\%$ proteins.
Here, $c$ is chosen by the investigator and should reflect the effort and capacity to investigate selected proteins further.
\cite{lin2006loss} cast the problem as a Bayesian decision problem.
Let $ R_{i} = \rank(\gamma_{i}) = \sum_{i^{\prime}=1}^{I} \bm{1}(|\gamma_{i}| \ge |\gamma_{i^{\prime}}|),$ denote the true ranks, with $R_{i}=1$ for the smallest $|\gamma_{i}|$ and $R_{i}=I$ for the largest $|\gamma_{i}|$.
Alternatively, we use $P_{i} = R_{i}/(I+1)$ to report the quantile.
\cite{lin2006loss} consider (among others) the following loss aimed at identifying the $c-$quantile.
Let $\hR_{i}$ denote the estimated rank for unit $i$, $\hP_{i} = \hR_{i}/(I+1)$, $\bP=(P_{1},\ldots,P_{I}), \bhP=(\hP_{1},\ldots, \hP_{I})$, and define
\begin{align*}
  &L_{c}(P,\hP) = \frac1I \{\text{\# misclassifications}\} =\\
  &\frac1I \left\{ \sum_{i=1}^{I} \hbox{AB}(c, P_{i}, \hP_{i}) + \hbox{BA}(c,P_{i}, \hP_{i}) \right\}
\end{align*}
where $\hbox{AB}(c, P,\hP)= \bm{1}(P>c, \hP <c) = \bm{1}(R > c(I+1), \hR < c(I+1))$ and $\hbox{BA}(c, P,\hP)= \bm{1}(P<c, \hP >c) = \bm{1}(R < c(I+1), \hR > c(I+1))$ count false positives and false negatives.
\cite{lin2006loss} show that $L_{c}$ is optimized by $\hR_{i}=\Rstar_{i}$ with
\begin{align}\label{eq: rank prot}
  \Rstar_{i}(c)= \rank \{p(P_{i}>c \mid \by) \}.
\end{align}
We implement inference under the proposed model using MCMC posterior simulation as described in Section S.4 of the supplementary materials.
Some summaries are shown in Figure~\ref{fig: prot profile}.
Figure \ref{fig: prot profile} (left) shows fitted protein expression profiles for two randomly chosen proteins, together with the observed data from patients $j=1,\ldots, J$, including controls and cases.

Figure \ref{fig: prot profile} (right) shows co-clustering probabilities for the random partition of proteins that arises from the PYP prior \eqref{PYP-DDP}.
The arrangement of ties among the $\xi_{i}$ defines protein clusters, and similarly, ties of $\eta_{j}$ define patient clusters.
The figure shows co-clustering probabilities for proteins.
Let $K_{I}$ and $K$ denote the (random) number of protein and subject clusters, respectively.
We find $p(21 \le K \le 28 \mid data) \approx 90\%$ for the number of patient clusters, with several patients in singleton clusters.
We find $p(K_{I}=20 \mid data) \approx 1$ for the random partition of proteins.
The latter is due to our choice of hyperparameters for the PYP prior with a negative discount parameter (see Section S.4.1 in the supplementary materials), which imposes an upper bound on $K_{I}$.
The upper bound on the number of clusters was needed, mainly to control computational effort.

Back to the main inference goal of identifying proteins related to ataxia.
Evaluating the posterior rank summary \eqref{eq: rank prot}, we find the top $100$ ($=2.5\%$) ranked proteins.
Table \ref{tab: top_gamma} shows the top $20$ of these.
\begin{table}[ht]
\centering
\footnotesize
\begin{tabular}{ccccc}
    \hline
    Q04917 & Q05329 & Q9Y371-2 & O60784 & P48728 \\
    \hline
    Q14974 & Q13616 & Q96A49 & P43155 & A0AVT1-1 \\ 
    \hline
    P54136-1 & P49327 & Q96RU3-1 & Q8NB37-2 & Q9P000-1 \\
    \hline
    Q9NSD9 & Q15813-2 & Q9Y4W6 & Q96PU8 & O75083\\
    \hline
\end{tabular}
\vspace{3mm}
\caption{Top 20 Ranked Proteins with highest posterior $|\gamma_{i}|$.}
\label{tab: top_gamma}
\end{table}

\section{Discussion}\label{sec: discussion}
We have argued for the use of separate exchangeability as a modeling principle, especially for nonparametric Bayesian models.
The main arguments are that (i) in several cases separate exchangeability more faithfully represents the experimental setup than, say, partial exchangeability; (ii) some summaries of interest (related to the identity of the experimental units) cannot even be stated without introducing separate exchangeability, and (iii) importantly, we illustrate how to develop tractable and flexible separately exchangeable counterparts of a particular BNP model that introduced partial exchangeability (plus exchangeability of groups) by means of setting up dependent random probability measures.
An example of the latter is inference on shared nested partitions of rows across different columns in a data matrix, instead of sharing the law as in NDP.
In a wider context, the discussion shows that careful consideration of the experimental setup often leads to more specific symmetry assumptions than omnibus exchangeability or partial exchangeability.

Finally, recall that the proposed separately exchangeable BNP models in both examples can be rephrased in terms of dependent random partition models that exploit the random partitions induced by the ties implied by sampling from discrete random measures.
We proposed natural generalizations to similar compositions of Gibbs-type priors \citep{gnedin2006exchangeable, deblasi2015gibbs} that preserve analytical and computational tractability of the random partition law that allows us to perform inference in practice, as illustrated.

\section*{Data and Code}
Data and code to reproduce the results in the manuscript are available at \url{https://github.com/GiovanniRebaudo/SEP-BNP}.

\section*{Acknowledgment}
The authors are grateful to the Editor, the AE, and three anonymous referees, whose feedback has led to a significant improvement in the manuscript.
Most of the paper was completed while G.\ R.\ was a Postdoc at UT Austin.
G.\ R.\ acknowledges support of MUR - Prin 2022 - Grant no.\ 2022CLTYP4, funded by the European Union - Next Generation EU.

\printbibliography
\newpage

\setcounter{equation}{0}
\setcounter{page}{1}
\setcounter{table}{0}
\setcounter{figure}{0}
\setcounter{section}{0}
\numberwithin{table}{section}
\renewcommand{\theequation}{S.\arabic{equation}}
\renewcommand{\thesubsection}{S.\arabic{section}.\arabic{subsection}}
\renewcommand{\thesection}{S.\arabic{section}}
\renewcommand{\thepage}{S.\arabic{page}}
\renewcommand{\thetable}{S.\arabic{table}}
\renewcommand{\thefigure}{S.\arabic{figure}}
\vspace{0cm}

\begin{center}
{\LARGE{Supplementary materials of \\
\bf		Separate Exchangeability as Modeling Principle in Bayesian Nonparametrics}}
\end{center}

\begin{center}
	\small
	\baselineskip=14pt
        Giovanni Rebaudo$^{a}$ (giovanni.rebaudo@unito.it) \\
        Qiaohui Lin$^{b}$ (qiaohui.lin@utexas.edu)\\
        Peter M\"uller$^{c}$ (pmueller@math.utexas.edu) \\
\vskip 3mm
$^{a}$ESOMAS Dept., University of Torino and Collegio Carlo Alberto, IT\\
\vskip 4pt 
$^{b}$Genentech, USA
\vskip 4pt 
$^{c}$SDS \& Mathematics Depts., University of Texas at Austin, USA\\ 
\end{center}

\newrefsection

\section{Diet Swap Study}
\label{sec: Microbiome}
\begin{figure}
\centering
\includegraphics[width= 0.49\textwidth]{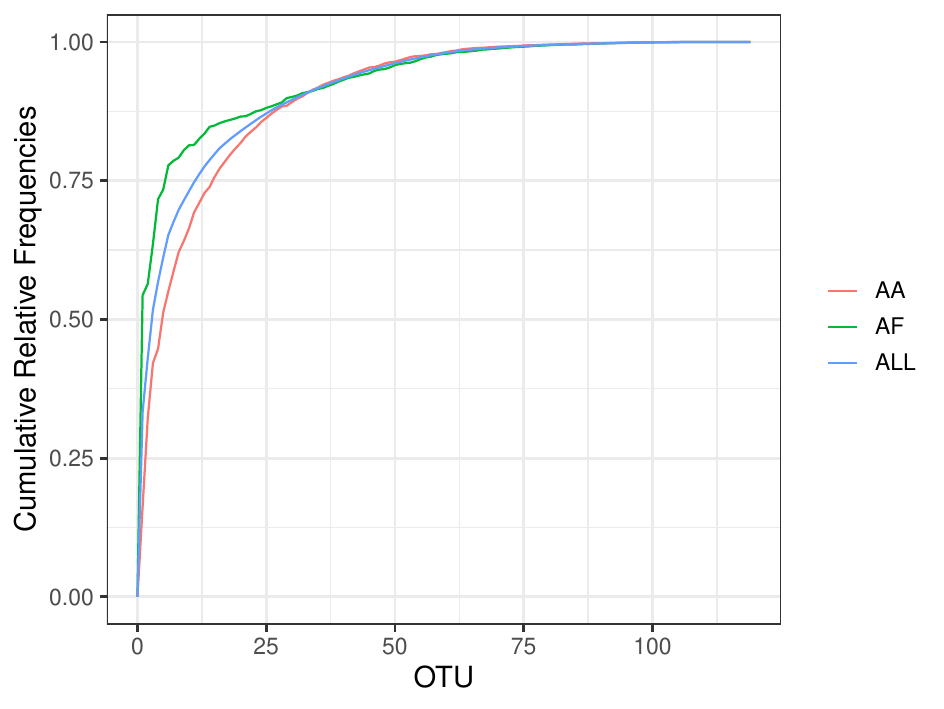}\\
\includegraphics[width= 0.49\textwidth]{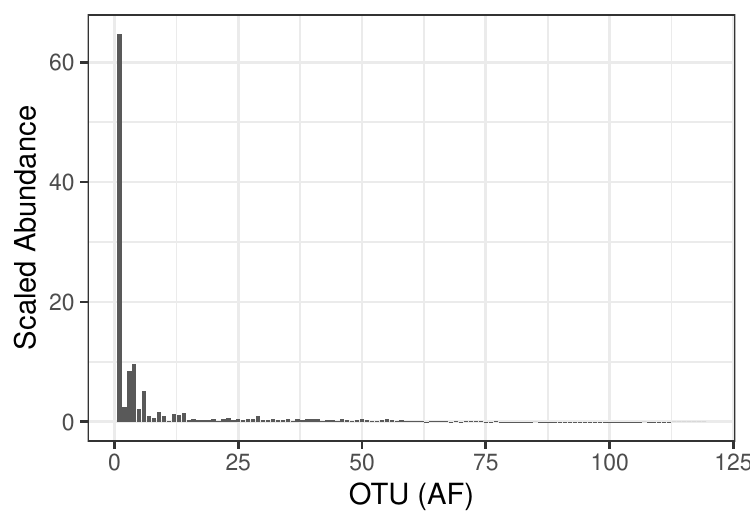} 
 \includegraphics[width= 0.49\textwidth]{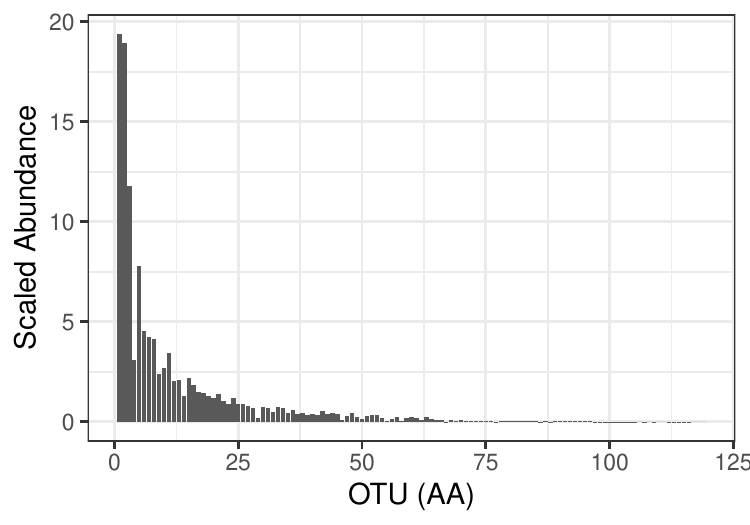}
\caption{Top: Cumulative relative frequencies of OTUs for average rural Africans (AF), average African Americans (AA), and the average of all subjects (ALL).
Bottom: Histogram of OTU abundance in AF and AA (scaled as described in the text).
}
\label{fig:summary-otu}
\end{figure}

We show some summary plots of the data to motivate the proposed inference.
First, we sort OTUs by overall abundance, i.e., distributions (across all subjects).
Figure~\ref{fig:summary-otu} (top) shows the cumulative relative frequencies of OTUs in rural Africans (AF), African Americans (AA), and all subjects, highlighting a difference in the distribution of OTU frequencies between AF and AA.
Throughout, the OTU frequencies in this data have been scaled by average library size, i.e., $y_{i,j}$ is the absolute count of OTU $z_{i,j}$ normalized by the totals $\gamma_{j}=\sum_{i=1}^{I} z_{i,j}$ multiplied by the average library size.
The two histograms at the bottom of the same figure show OTU abundance in the two groups, suggesting that subjects might meaningfully group by the distribution of OTU frequencies.

\begin{figure}[t]
    \centering
    \includegraphics[clip,width=0.6\textwidth]{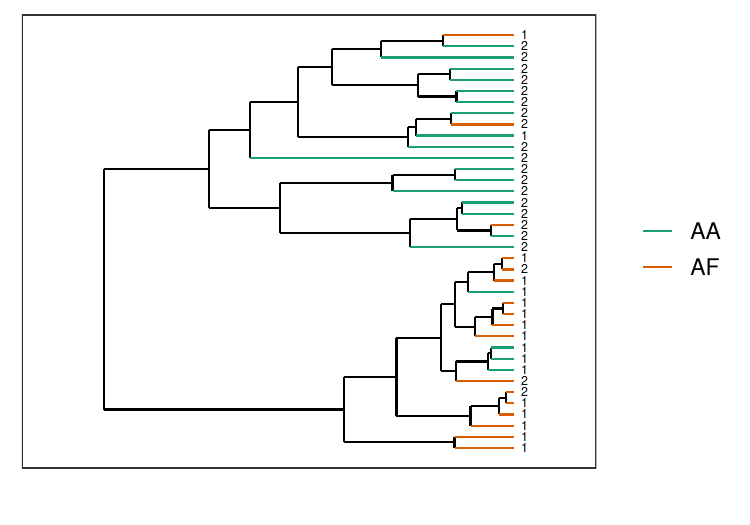}
    \caption{Agglomerative hierarchical clustering of subjects
      (colored according to the groups, AF and AA) with Euclidean
      distance and complete method using the $10$ OTUs with the highest cross-subject variance.}
    \label{fig:dend-subj-otu}
\end{figure}

In Figure~\ref{fig:dend-subj-otu}, we show (deterministic) hierarchical clustering of the subjects based on the $10$ OTUs with the highest empirical variance.
Note how the clusters correlate well with the two groups, AF and AA, suggesting that in grouping subjects by distributions of OTU frequencies, we should proceed in a way that maintains and respects OTU identities (as is the case in the hierarchical clustering).
We wish to do so in a way that can still account for substantial uncertainty (where uncertainty is meant around the point estimate of the clustering structure, no matter if it is interpreted as an estimator of an unknown truth or just as a summary of the data).
Observing these features in the figures motivates us to formalize inference on grouping subjects by OTU abundances using model-based inference.
We will set up a separately exchangeable model that allows us to respect both subject and OTU indices.

\begin{figure}
  \centering
  \includegraphics[width=0.4\textwidth]{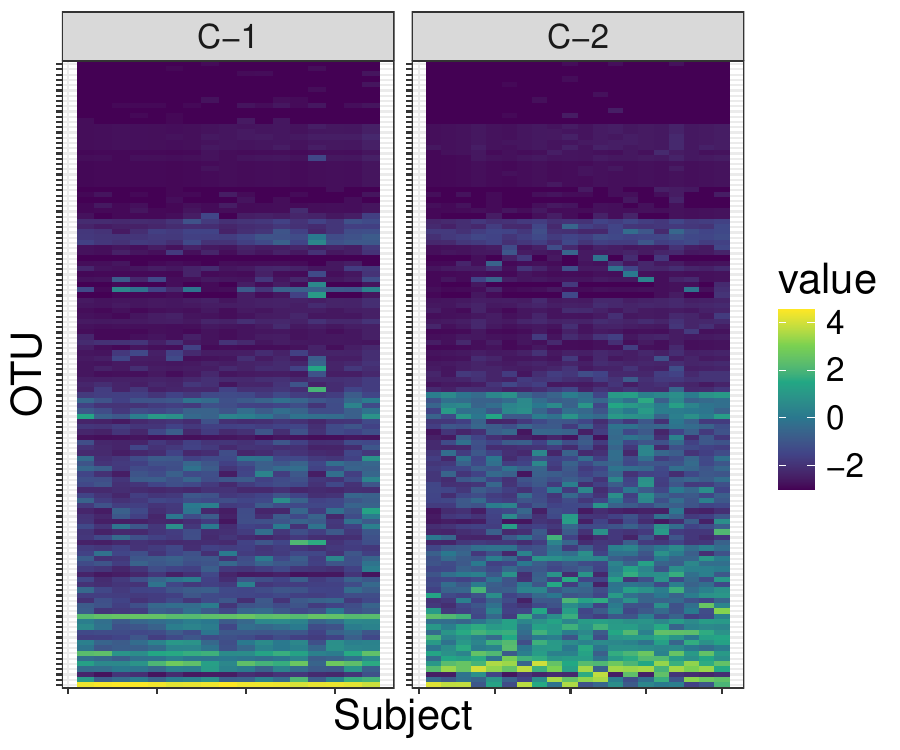}
  \caption{Heatmap of column-scaled $y$ in log scale for each cluster of subjects.
  OTUs are sorted by each cluster-specific OTU cluster assignment.}
  \label{fig: mb-heatmap-y}
\end{figure}

Figure~\ref{fig: mb-heatmap-y} shows the same nested partitions as in Figure 3, 
but now by showing the data $y_{i,j}$ arranged by subject clusters.
In each panel, OTUs are sorted by observational clusters.
That is, each plot shows the data corresponding to $j \in C_{k}$, for $k=1$ and $2$ (the singleton cluster $k=3$ is not shown).
The subjects $j \in C_{k}$ are on the x-axis.
The y-axis plots OTUs, arranged by the estimated observational clusters.
The patterns in $y_{i,j}$ echo the clusters shown in the previous plot.
Inference as in Figure 3 
or \ref{fig: mb-heatmap-y} is not meaningfully possible under partial exchangeability, since the nested partitions $\Psi_{j}=(R_{j,\ell}, \ell=1,\ldots, L)$ are not shared across $j \in C_{k}$ \mbox{(only the laws of $\Psi_{j}$ are shared)}.

\section{Proof}
\paragraph*{\em Claim:} 
Let $\mathbf{X}$ be an infinite partially exchangeable array (4) of the main paper 
with $x_{i,j}$ real-valued, square-integrable random variables such that the correlations are well-defined, and the marginal correlations within columns are equal, i.e., 
$
\corr(x_{i,j},x_{i^{\prime}, j}) = \corr(x_{i,j^{\prime}},x_{i^{\prime}, j^{\prime}}).
$
Thus,
\begin{align}\label{appxA}
  \corr(x_{i,j},x_{i^{\prime}, j}) \ge \corr(x_{i,j},x_{i^{\prime}, j^{\prime}}) \quad j \ne j^{\prime}, \,\ i \ne i^{\prime}
\end{align}
and strict inequality can occur.
\begin{proof}
We define $U_{i,j} \coloneqq \mbox{Var}(x_{i,j})^{-1/2} x_{i,j}$ such that $\cov(U_{i,j},U_{i^{\prime}, j})=\corr(x_{i,j},x_{i^{\prime}, j})$.
By de Finetti's Theorem (4) 
and the law of total covariance
    \begin{align*}
    &\cov(U_{i,j},U_{i^{\prime}, j}) = 0 +
    \cov[\EV(U_{i,j}\mid P_{j}), \EV(U_{i^{\prime}, j} \mid P_{j})] = \var[\mathbb{E}(U_{i^{\prime}, j} \mid P_{j})]\\
        & \ge 0 + \cov[\EV(U_{i,j}\mid P_{j}), \EV(U_{i^{\prime}, j^{\prime}} \mid P_{j^{\prime}})]
        = \cov(U_{i,j},U_{i^{\prime}, j^{\prime}}),
    \end{align*}
where the $\ge$ holds by Cauchy-Schwarz inequality after noting that $\var[\mathbb{E}(U_{i^{\prime}, j} \mid P_{j})] = \corr(x_{i,j}, x_{i^{\prime},j}) =   \corr(x_{i,j^{\prime}}, x_{i^{\prime},j^{\prime}}) =  \var[\mathbb{E}(U_{i^{\prime}, j^{\prime}} \mid P_{j^{\prime}})]$.
Considering a trivial example with $>$ in \eqref{appxA} proves the claim.
\end{proof}

\section{Posterior inference under a separately exchangeable prior for nested partitions}\label{sec: MCMC sep part} 

First, recall the joint probability model in (15) of the main manuscript. 
In practice, we can use an implementation using a finite DP approximation \citep{ishwaran2001gibbs} for $G_{k}$, i.e., we truncate $G_{k}$ with $L$ atoms.
Similarly, we can truncate the stick-breaking prior for $\pi_{k}$ at a fixed number of $K$ atoms (with $\pi_{K} = 1-\sum_{k=1}^{K-1} \pi_{k}$).
Let then $\bpi=(\pi_{1},\ldots,\pi_{K})$ and $\bw_{k} = (w_{k,1},\ldots,w_{k,L})$.
We implement MCMC simulation to generate from (15) under these finite DP approximations:
\begin{align}
\begin{split}
    &p(\bS, \bM, \bpi, \bw, \bmu, \bSigma \mid y_{1:I,1:J}, K,L) \propto \prod_{j=1}^{J} \prod_{i=1}^{I} p(y_{i,j} \mid \mu_{M_{i,S_{j}}},\sigma^{2}_{M_{i,S_{j}}})\\
    & \quad \quad \times 
    \prod_{j=1}^{J} \pi_{S_{j}} \left\{ \prod_{k=1}^{K} \prod_{i=1}^{I} w_{k,M_{i,k}} \right\} \;
    p(\bpi)\, \prod_{k=1}^{K} p(\bw_{k})\, \prod_{\ell=1}^{L}
    p(\mu_{\ell}, \sigma_{\ell}^{2}),
\label{eq: sep rpm joint trunc}
\end{split}
\end{align}
with $p(\bpi) = p(\pi_{1},\ldots,\pi_{K})=\text{TGEM}(\beta)$ and $p(\bw_{k}) = p(w_{k,1}, \ldots, w_{k,L}) = \text{TGEM}(\alpha)$ being truncated stick breaking priors that approximate the $\GEM(\beta)$ and $\GEM(\alpha)$, respectively.  
Transition probabilities are defined by the following complete conditional posterior distributions in \eqref{eq: sep rpm joint trunc}.
We write $\bm{\theta}_{\ell}$ for $\bm{\theta}_{\ell}=(\mu_{\ell},
\sigma_{\ell}^{2})$ and
$C_{k}=\{j:\, S_{j}=k\}$ for the subject clusters.
Also, let $C_{k}^{-(j)} = C_{k} \setminus \{j\}$ and $C_{k}^{+(j)} = C_{k} \cup \{j\}$, i.e., $C_{k}$ without and with subject $j$ under consideration, respectively.
Below we use the notation $N(x \mid m, V)$ for an $N(m, V)$ probability density function (p.d.f.) evaluated at $x$, and we suppress $x$ if the argument is understood from the context.
\\
\\
\textbf{Algorithm 1:} MCMC algorithm for posterior inference under separately exchangeable NDP-CAM.
We define transition probabilities to update parameters in the following sequence.
\begin{enumerate}
\item $S_{j}$, $j=1,\ldots,J$:\\
The complete conditional posterior probabilities are $p(S_{j} = k \mid \bM, \ldots) \propto \prod_{i=1}^{I} p(y_{i,j} \mid \bm{\theta}_{M_{i,k}}) \cdot \pi_{k}$.
However, conditioning on the nested partition $\bM$ leads to a slowly mixing Markov chain.
We consider instead marginal probabilities, after integrating out $M_{i,k}$.
Note that marginalizing with respect to $M_{i,k}$ makes $(y_{i,j^{\prime}}: \, j^{\prime} \in C_{k})$ dependent.
Let $\bm{y}^{\star}_{i,k}=(y_{i,j}: \, j \in C_{k})$, $\bm{y}^{\star+(j)}_{i,k}=(y_{i,j^{\prime}}: \, j^{\prime} \in C_{k}^{+(j)})$ and $\bm{y}^{\star-(j)}_{i,k}=(y_{i,j^{\prime}}: \, j^{\prime} \in C_{k}^{-(j)})$ denote $y_{i,j}$ arranged by cluster (with or without subject $j$).
We get, now without conditioning on $\bM$:
\begin{align}
\begin{split}
    p(S_{j} = k \mid \bm{\theta}, \bpi, \bw, \by, \bS_{-j}) &\propto \pi_{k} \times   \prod_{i=1}^{I} p(y_{i,j} \mid \bm{y}^{\star-(j)}_{i,k}, \bm{\theta}, \bw_{k})\\
    &= \pi_{k} \times \prod_{i=1}^{I} \frac{\sum_{\ell=1}^{L} w_{k,\ell}\, p(\bm{y}^{\star+(j)}_{i,k} \mid \bm{\theta}_{\ell})} {\sum_{\ell=1}^{L} w_{k,\ell}\, p(\bm{y}_{i,k}^{\star-(j)} \mid \bm{\theta}_{\ell})}
    \label{pSj}
\end{split}
\end{align}
$k=1,\ldots, K$.
Using \eqref{pSj}, we implement an MH transition probability for $S_{j}$.
As a proposal distribution, we use the marginal probability $p(S_{j}=k \mid \bm{\theta},\bpi,\bw, \by_{j})$ (conditional on $\by_{j}$ only).
\item $M_{i,k}$, $i=1,\ldots,I$, $k=1,\ldots,K$:
$$
  p(M_{i,k} = \ell \mid \ldots) \propto \prod_{j: S_{j}=k} p(y_{i,j} \mid \mu_{\ell},\sigma_{\ell}^{2}) \times w_{k,\ell}; \quad \ell=1,\ldots, L.
$$
\item $\mu_{\ell}$, $\ell=1,\ldots,L$:
Let $C_{\ell}=\{(i,j):\, M_{i,S_{j}}=\ell\}$ and $n_{\ell}=|C_{\ell}|$,
$$
    p(\mu_{\ell} \mid \ldots) \propto \N(\mu_{\ell} \mid \mu_{0},\sigma_{0}^{2}) \times \prod_{(i,j) \in C_{\ell}} p(y_{i,j} \mid \mu_{\ell},\sigma_{\ell}^{2}) \propto \N(m,V)
$$
with $1/V = 1/\sigma_{0}^{2}+n_{\ell}/\sigma_{\ell}^{2}$ and $m=V(\mu_{0}/\sigma_{0}^{2}+\sum_{C_{\ell}} y_{i,j}/\sigma_{\ell}^{2})$.
\item $\sigma_{\ell}^{2}$, $\ell=1,\ldots,L$:
$$
    p(\sigma_{\ell}^{2} \mid \ldots) \propto \IG(\sigma_{\ell}^{2} \mid a,b) \times \prod_{(i,j) \in C_{\ell}} p(y_{i,j} \mid \mu_{\ell},\sigma_{\ell}^{2}) \propto \IG(a_{\ell},b_{\ell})
$$
with $a_{\ell}=a+n_{\ell}/2$ and $b_{\ell}=b+\sum_{C_{\ell}} (y_{i,j}-\mu_{\ell})^{2}/2$.
\item $w_{k,\ell}$: let $w_{k,\ell} = v_{k,\ell}\, \prod_{h<\ell}(1-v_{k,h})$.
Let $A_{k,\ell}=\{(k,i): \, M_{i,k}=\ell\}$, $B_{k,\ell}=\{(k,i): \, M_{i,k}>\ell\} = \bigcup_{h>\ell} A_{k,h}$, and $n_{k,\ell}=|A_{k,\ell}|$.

Then, for $k=1,\ldots,K$, $v_{k,L}= 1$ and for $\ell=1,\ldots,L-1$:
\begin{align*}
  p(v_{k,\ell} \mid \ldots) &\propto \text{Beta}(v_{k,\ell} \mid 1,\alpha) \times \prod_{(i,k) \in A_{k,\ell}} v_{k,\ell}\times \prod_{(i,k) \in B_{k,\ell}} (1-v_{k,\ell}) \\
  &\propto \text{Beta}(1+n_{k,\ell}, \alpha + \sum_{h>\ell} n_{k,h}).
  \end{align*}
\item $\pi_{k}$: let $\pi_{k} = v_{k} \prod_{h<k}(1-v_{h})$, $C_{k}=\{j: \, S_{j}=k\}$, $D_{k} = \bigcup_{h>k} C_{h}$ and $n_{k} = |C_{k}|$ (as before).
Then $v_{K}=1$ and for $k=1, \ldots, K-1$:
$$
  p(v_{k} \mid \ldots) \propto \text{Beta}(v_{k}\mid 1,\beta) \times \prod_{j \in C_{k}} v_{k}\times \prod_{j \in D_{k}} (1-v_{k}) \propto \text{Beta}(1+n_{k}, \beta + \sum_{h>k} n_{h}).
$$
\end{enumerate}

\section{Implementing inference under separately exchangeable BNP regression}\label{sec: alg 2}
We are mainly interested in identifying the proteins with the highest rank in $|\gamma_{i}|$ (the difference in slope between patients and controls).
The latter proteins are the ones that are most likely linked with ataxia.

\subsection{Hyperparameter settings}
\label{sec: hyper}
In the analysis, we set the following hyperparameters.
First, we set the $a_{\sigma}$ and $b_{\sigma}$ such that the prior mean and variance of $\sigma^{2}$ are both equal to $1$.
Then, we set $\alpha_{\eta}=0.1, \sigma_{\eta}=0.1, \alpha_{\xi}=1$, and $\sigma_{\xi}=-1/20$.
The negative discount parameter $\sigma_{\xi}<0$ implies an upper bound on the number of protein clusters: $\alpha_{\xi}/|\sigma_{\xi}|=20$.
Finally, we set $\mu_{\eta}=\bar{y} = \frac{\sum_{i,j,x}y_{i,j}(x)}{N_{tot}}$, $N_{tot} = \sum_{i,j,x} \bm{1}(y_{i,j}(x) \text{ is observed})$, $\Sigma_{\eta}=25, \mu_{\xi}=(0, \ldots, 0)^{\top} \in \R^{12} $, and $\Sigma_{\xi}=I_{12}$.
The large variance $\Sigma_{\eta}$ implements a vague hyperprior, while $\Sigma_{\xi}=I$ imposes moderate prior shrinkage to a flat (over time) protein expression profile.

\subsection{Characterization of the model as a composition of gCRPs}
To derive the MCMC algorithm for posterior inference under separately exchangeable BNP regression, we recall the model.
\begin{align}\label{eq: sep reg prot supp}
\begin{split}
    &y_{i,j}(X) \mid \xi_{i}, \eta_{j}, \sigma^{2} \sim \N_{2T}(\eta_{j} + \bD_{\xi} \xi_{i}, \, \sigma^{2} I_{2T}), \quad \sigma^{2} \sim \text{Inv-Ga}(a_{\sigma}, b_{\sigma}) \\
    &\eta_{j} \mid P_{\eta} \simiid P_{\eta}, \quad P_{\eta} \sim \text{PYP}(\alpha_{\eta}, \sigma_{\eta}, \N_{1}(\mu_{\eta}, \Sigma_{\eta})),\\
    &\xi_{i} \mid P_{\xi} \simiid P_{\xi}, \quad P_{\xi} \sim \text{PYP}(\alpha_{\xi}, \sigma_{\xi}, \N_{12}(\mu_{\xi}, \Sigma_{\xi})).
\end{split}
\end{align}
Note that with a sum of a scalar and a vector (or a matrix), we refer to the addition of the scalar to each element of the vector (or the matrix).
For the statement of the detailed transition probabilities in Algorithm 2, it is useful to characterize the mixture model \eqref{eq: sep reg prot supp} with an equivalent hierarchical model that considers the clustering assignment indicators that follow the gCRP induced by the PYP.
\begin{align*}
    &y_{i,j}(X) \mid (\tilde{\xi}_{h}), (\tilde{\eta}_{s}), s_{i,\xi}, s_{j,\eta}, \sigma^{2} \sim \N_{2 T}(\tilde{\eta}_{s_{j,\eta}} + \bD_{\xi} \tilde{\xi}_{s_{i,\xi}}, \, \sigma^{2} I_{2 T}), \quad \sigma^{2} \sim \text{Inv-Ga}(a_{\sigma}, b_{\sigma}) \\
    &\tilde{\eta}_{s} \simiid \N_{1}(\mu_{\eta}, \Sigma_{\eta}), \quad 
    (s_{j,\eta})_{j=1}^{\infty} \sim \text{gCRP}(\alpha_{\eta}, \sigma_{\eta}),\\
    &\tilde{\xi}_{h} \simiid \N_{12}(\mu_{\xi}, \Sigma_{\xi}), \quad 
    (s_{i,\xi})_{i=1}^{\infty} \sim \text{gCRP}(\alpha_{\xi}, \sigma_{\xi}).
\end{align*}
Here the $\text{gCRP}(\alpha, \sigma)$ denotes the generalized Chinese restaurant process induced by the $\text{PYP}(\alpha, \sigma; \cdot)$ (compare Table 1), 
and $\teta_{s}$ and $\txi_{h}$ are the unique atoms (in order of arrival) of the random mixing measures $P_{\eta}$ and $P_{\xi}$ in \eqref{eq: sep reg prot supp}, respectively.
Interpreting $s_{j,\eta}$ and $s_{i,\xi}$ as cluster membership indicators defines random partitions of subjects and proteins, respectively.
Let $C_{h,\eta}=\{j:\, s_{j,\eta}=h\}$ denote the subject cluster $h$ defined by $s_{j,\eta}$, and let $n_{h,\eta}=|C_{h,\eta}|$ its frequency for $h\in[K]$, and similar for $C_{h,\xi}$ and $n_{h,\xi}$.

\subsection{Marginal Gibbs sampler}
We set up a posterior MCMC simulation for posterior inference under separately exchangeable BNP regression.
We describe the implementation for the protein profile example from Section 5.2 of the main manuscript.
Let $\ty_{i,j}=y_{i,j}(\bx_{j})$ denote the observed protein expression of protein $i$ for subject $j$ with covariates $\bx_{j} = (z_{j},t_{j})$ including condition $z_{j} \in \{0,1\}$ and age $t_{j}$, and let $\bd_{j}=\bd_{\xi}(\bx_{j})$ denote the corresponding design vector for subject $j$.
\\
\\
\textbf{Algorithm 2:} MCMC algorithm for posterior inference under separately exchangeable BNP regression.
Steps 1-5 define posterior simulation for the example in Section 5.2 
with $D_{\beta}=\emptyset$ and $D_{\eta}=1$, and with a single observation $\ty_{i,j} = \by_{i,j}(\bx_{j})$ for each $(i,j)$ in (24). 
With suitable modifications, similar types of transition probabilities can be defined for a more general instance of the model with non-zero $D_{\beta}$ and different $D_{\eta}$.
\begin{enumerate}
  \item
    For each subject $j \in [J]$, sample $s_{j,\eta}$ from $p(s_{j,\eta} =s \mid \cdots)$ defined as follows.
    Let $\yd_{i,j} = \ty_{i,j}- \bd_{j} \txi_{s_{i,\xi}}$.
    Then
    \begin{equation}\label{eq:pseta}
      p(s_{j,\eta} =s \mid \cdots) \propto
      \begin{cases}
        W_{J,K^{-j}}
        (n_{s, \eta}^{-j}-\sigma_{\eta}) \, \prod_{i} \N(\yd_{i,j} \mid \teta_{s}, \sigma^{2}) &
        s \in [K^{-j}] 
        \vspace*{0.2cm}\\
        W_{J,K^{-j}+1} \, 
        g_{\text{new},\eta}(\yd_{i,j}) & s = K^{-j} +1.
      \end{cases}
    \end{equation}
Here and in the next equation the product over $i$ goes over all proteins with observed $\ty_{i,j}$, and 
\begin{align}
\begin{split}
\label{gnew}
  g_{\text{new}, \eta}(\yd_{i,j})
  &= \int \prod_{i} \N_{1}(\yd_{i,j} \mid \teta, \sigma^{2})
    \, \N_{1}(\teta \mid \mu_{\eta}, \Sigma_{\eta})\, \d\teta\\
    &= \N_{I}(y_{1:I,j}^{\dagger} \mid \mu_{\eta} \bone, \sigma^{2} I + \Sigma_{\eta} \bone\bone^{\top}\big).
\end{split}
\end{align}
In the implementation, $g_{\text{new}, \eta}(\cdot)$ is best evaluated using the candidate formula \citep{besag1989candidate} and note that the dimension of the multivariate density $\N_{I}$ can actually be smaller than $I$ in case of missing values.
Also in the implementation, for $n_{s, \eta}^{-j}<n^{\star}$ we analytically marginalize \eqref{eq:pseta} with respect to $\teta_{s}$.
We use $n^{\star}=20$.
\item For each subject-cluster $s$, sample $\teta_{s}$ from $p(\teta_{s}\mid \cdots)=N(m_{s},V_{s})$ with the moments determined as follows.
Let $\yd_{i,j}$ be defined as in the previous step.
Let $A_{s} = \{(i,j):\, s_{j,\eta}=s, \ty_{i,j} \text{ is observed}\}$.
Then $(m_{h}, V_{h})$ are defined as the posterior moments under the conjugate normal/normal model with a normal sampling of $\yd_{i,j} \sim \N(\teta_{s},\sigma^{2})$, $(i,j) \in A_{s}$ and the conjugate normal prior, $\teta_{s} \sim \N(\mu_{\eta}, \Sigma_{\eta})$.
\item For each protein $i \in [I]$, sample $s_{i,\xi}$ from $p(s_{i,\xi}=h\mid \ldots)$ defined as follows.
Let $\yd_{i,j} = \ty_{i,j}- \teta_{s_{j,\eta}}$.
Then
\begin{equation*} p(s_{i,\xi}=h\mid \ldots) \propto
\begin{cases} W_{I,K_{I}^{-i}} (n_{h, \xi}^{-i}-\sigma_{\xi}) \, \prod_{j} \N(\yd_{i,j} \mid \bd_{j} \txi_{h},\sigma^{2}) & h \in [K_{I}^{-i}] \vspace*{0.2cm}\\ W_{I,K_{I}^{-i}+1} \, g_{\text{new},\xi}(\yd_{i,j}) & h = K_{I}^{-i} +1.
\end{cases}
\end{equation*}
Here and in the next equation the product over $j$ goes over all subjects $j$ with observed $\ty_{i,j}$, and similar to \eqref{gnew}, 
\begin{align*}
    g_{\text{new}, \xi}(\yd_{i,j}) &= \int \prod_{j} \N(\yd_{i,j} \mid \bd_{j} \txi, \sigma^{2}) \, \N_{12}( \txi \mid \mu_{\xi}, \Sigma_{\xi})\, \d\txi\\
    & = \N_{J} \big(y_{i,1:J}^{\dagger} \mid \bD\mu_\xi,\ \sigma^{2} I + \bD\Sigma_\xi \bD^{\top}\big),
\end{align*}
where $\bD$ stacks the row vectors $\bd_{j}^{\top}$ and note the dimension of the multivariate density $\N_{J}$ can be actually smaller than $J$ in case of missing values.
In the implementation, $g_{\text{new}, \xi}(\cdot)$ is best evaluated using the candidate formula.
Also, similar to Step 1, for $n_{h, \xi}^{-i}<n^{\star}$ we marginalize analytically with respect to $\txi_{h}$.

\item For each protein-cluster $h$, sample $\txi_{h} \sim \N(m_{h}, V_{h})$ with the moments determined by setting up a normal linear regression model.
Alternatively, in the implementation, we can use a random scan Gibbs sampler, updating $s_{i,\xi}$ for a randomly chosen protein.
Let $\yd_{i,j}$ be defined as in step 3, and $B_{h} = \{(i,j):\, s_{i,\xi}=h, \ty_{i,j} \text{ is observed}\}$.
Consider then a normal linear regression with $\yd_{i,j} \sim \N(\bd_{j} \txi_{h}, \sigma^{2})$, $(i,j) \in B_{h}$ and conjugate prior $\txi_{h} \sim \N(\mu_{\xi}, \Sigma_{\xi})$.
The moments $(m_{h}, V_{h})$ are the posterior moments for $\txi_{h}$ in this regression problem.

\item Sample $\sigma^{2}$ from the complete conditional posterior.
Let $\yd_{i,j} = \ty_{i,j}-\teta_{s_{j,\eta}}-\bd_{j} \txi_{s_{i,\xi}}$, $D=\{(i,j):\, \ty_{i,j} \mbox{ is observed}\}$, and $N=|D|$.
Then
\begin{align*} p(\sigma^{2}\mid \cdots) =\IG\left(a_{\sigma} + N/2,
b_{\sigma}+ \frac{1}{2} \sum_{D} y^{\dagger 2}_{i,j} \right).
\end{align*}
\end{enumerate}

\section{Point estimates of multilevel partitions}
The choice of good summaries (i.e., point estimates) for reporting posterior inference on a random partition $\bm{C}$ is an important and nontrivial question in Bayesian analysis. 
A common decision-theoretic approach is to obtain the point estimates $\hat\Psi$ by minimizing the Bayesian risk, i.e., the expected value of a loss function with respect to the posterior distribution of the partition (often estimated with the empirical distribution over the MCMC samples after the burn-in): 
\begin{equation}\label{eq: optimal partition}
    \hat{\bm{C}} = \underset{\bm{C}^{\prime}}{\text{argmin}} \, \mathbb{E}[L(\bm{C}^{\prime}, \bm{C}) \mid \text{data}] = \underset{\bm{C}^{\prime}}{\text{argmin}} \, \sum L(\bm{C}^{\prime},\bm{C}) \, \mathbb{P}(\bm{C} \mid \text{data}). 
\end{equation}
For the loss function $L$, when estimating the partition of the columns $\bm{C}:=\{C_{1}, \ldots, C_{K}\}$, we consider the variation of information loss (VI) \citep{meila2007comparing} as suggested by \cite{wade2018bayesian} and implemented in \texttt{salso} \citep{dahl2022search,dahl2022salso}. 
Alternative loss functions can be used as required for specific applications, such as the Binder loss (BI) \citep{binder1978bayesian} or an entropy-regularized version of VI or BI losses \citep{franzolini2024entropy} that can enhance interpretability.

However, in the more challenging case of separately exchangeable nested partitions, we need two- (or more-) level point estimates for related partitions ($\bm{C}$ and $\Psi_{1}, \ldots, \Psi_{J}$) and it is desirable to have point estimates that also preserve the nested structures.
More precisely, we want a joint point estimate that is in the support of the multilevel partition.
In our example, the partitions of the OTUs should be the same for the subjects that are clustered together in the point estimates, as is the case in each (MCMC) joint realization.
That is, the report should respect
\begin{equation}\label{eq: ex nested constraint}
\Psi_{j}=\Psi_{j^{\prime}} \mid \{ \{j, j^{\prime}\} \subset C_{k} \text{ for some } k \in [K]\}.
\end{equation}
We use a simple technique that can be applied to two- or multi-level partitions in several dependence scenarios, also beyond separate exchangeability, such as under exchangeable models \citep[e.g.,][]{rebaudo2025graph} or under partially exchangeable models \citep[e.g.,][]{rodriguez2008nested, denti2023common, lijoi2023flexible}. 
The idea is to condition on the point estimates of the first level partition, such as for us the partition of the columns (i.e., the subjects), and given that estimate, consider the conditional distribution of the partitions in the other level, that is, the rows (i.e., the OTUs). 
That is, for each $j$
\begin{equation*}
\hat{\bm{\Psi}}_{j} = \underset{\bm{\Psi}_{j}^{\prime}}{\text{argmin}} \, \mathbb{E}[L(\bm{\Psi}_{j}^{\prime},\bm{\Psi}_{j}) \mid \text{data}, \hat{\bm{C}}] = \underset{\bm{\Psi}_{j}^{\prime}}{\text{argmin}} \, \sum L(\bm{\Psi}_{j}^{\prime},\bm{\Psi}_{j}) \, \mathbb{P}(\bm{\Psi}_{j} \mid \text{data}, \hat{\bm{C}}).
\end{equation*}
Note that this is simply implemented by estimating the first-level partition with an initial MCMC sample (after the burn-in), minimizing \eqref{eq: optimal partition} and then fixing the point estimates of the first-level partition for the remaining MCMC iterations that are used to estimate the second-level partition. 
By construction, the resulting multilevel point estimate respects 
\eqref{eq: ex nested constraint}.

\section{Software, Runtime, etc.}\label{sec: software supp}
Data and code to reproduce the results in the manuscript are available at \url{https://github.com/GiovanniRebaudo/SEP-BNP}.
We programmed everything in \texttt{R}.
The analyses are performed with a MacBook Pro 2023 with 16 GB of RAM (macOS Ventura), using \texttt{R} version 4.3.2.

\subsection{Microbiome}
\begin{figure}
    \centering
    \includegraphics[width=1\linewidth]{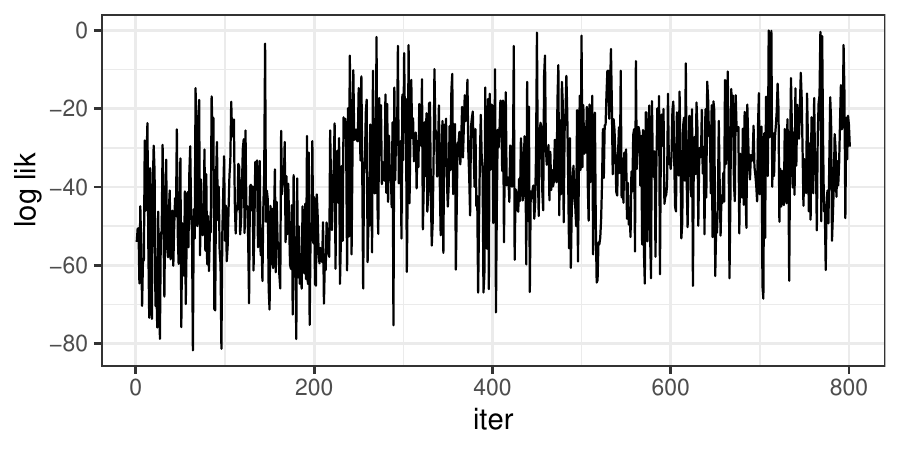}
    \caption{Microbiome Data: Trace-plots of the log-likelihood (up to a common shift) evaluated in the sampled parameters across MCMC iterations after burn-in and thinning.}
    \label{fig: trace ll rpm}
\end{figure}

The results reported in this article are based on $10,000$ MCMC iterations with the initial $2,000$ iterations discarded as burn-in and a thinning by batches of $10$. 
The MCMC algorithm takes less than $10$ minutes.
Figure \ref{fig: trace ll rpm} shows the trace plots of the log-likelihood evaluated in the MCMC iterations.
It does not suggest a convergence issue.

\subsection{Protein expression profiles}
\begin{figure}
    \centering
    \includegraphics[width=1\linewidth]{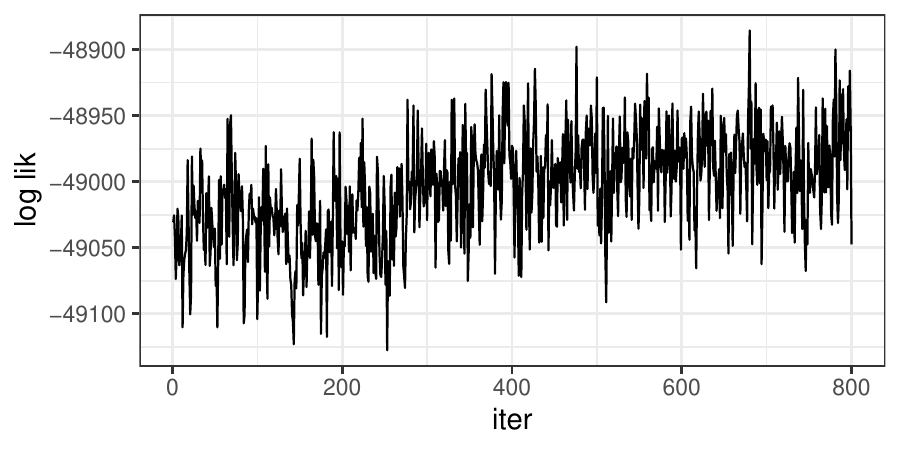}
    \caption{Protein Data: Trace-plots of the log-likelihood evaluated in the sampled parameters across MCMC iterations after burn-in and thinning.}
    \label{fig: trace ll reg}
\end{figure}

The results reported in this article are based on $10,000$ MCMC iterations with the initial $2,000$ iterations discarded as burn-in and a thinning in batches of $10$. 
The MCMC algorithm takes less than $1$ hour and $5$ minutes.
A trace plot of the log-likelihood evaluations is shown in Figure \ref{fig: trace ll reg}.
There is no evidence of convergence issues.

\section{Cubic B-splines}\label{sec: b-splines}
In the separately exchangeable BNP regression, we employed a cubic (i.e., $4$th-order) B-spline basis.
The construction of M-order B-spline bases \citep{boor1978practical, eilers1996flexible} is detailed below, and it is based on the slightly more general description in \cite{hastie2009elements}.

Let $\xi_{0}< \xi_{1} < \ldots < \xi_{K} < \xi_{K+1}$ be a sequence of knots that defines the domain where we will evaluate our spline and partition the interval $[\xi_{0}, \xi_{K+1}]$ in $K+1$ intervals.

We now consider the augmented knots
\begin{align*}
    \tau_{1} =& \ldots = \tau_{M} = \xi_{0} < \tau_{M+1}=\xi_{1} < \tau_{M+2}=\xi_{2} < \ldots \\
    &\ldots < \tau_{K+M}=\xi_{K} < \tau_{K+M+1}=\ldots = \tau_{K+2M}=\xi_{K+1}.
\end{align*}
Let $B_{i,m}(x)$ denote the $i$th B-spline basis function of order $m \le M$ for the augmented knot sequence.
The following recursive formula defines them.
For $i=1, \ldots, K+2M-1$.
\[
B_{i,1}(x) = \begin{cases}
    1 & \text{ if } \tau_{i} \le x < \tau_{i+1}\\
    0 & \text{ otherwise }
\end{cases}
\]
and for $i=1, \ldots, K+2M-m$
\[
    B_{i, m}(x) = \frac{x-\tau_{i}}{\tau_{i+m-1}-\tau_{i}} B_{i, m-1}(x) + \frac{\tau_{i+m}-x}{\tau_{i+m}-\tau_{i+1}} B_{i+1, m-1}(x),
\]
with the usual convention $0/0:=0$.
In the analysis, we consider $M=4$ (i.e., cubic B-spline basis functions), and $K=2$ internal knots (the $33\%$ and $66\%$ percentiles of the empirical distribution of the covariate).
Therefore, we obtain $B_{i,4}$, $i=1, \ldots, 6$ cubic B-spline basis functions for the knots $\xi_{0}< \xi_{1} < \ldots < \xi_{K} < \xi_{K+1}$.

\begin{figure}
    \centering
    \includegraphics[width=1\linewidth]{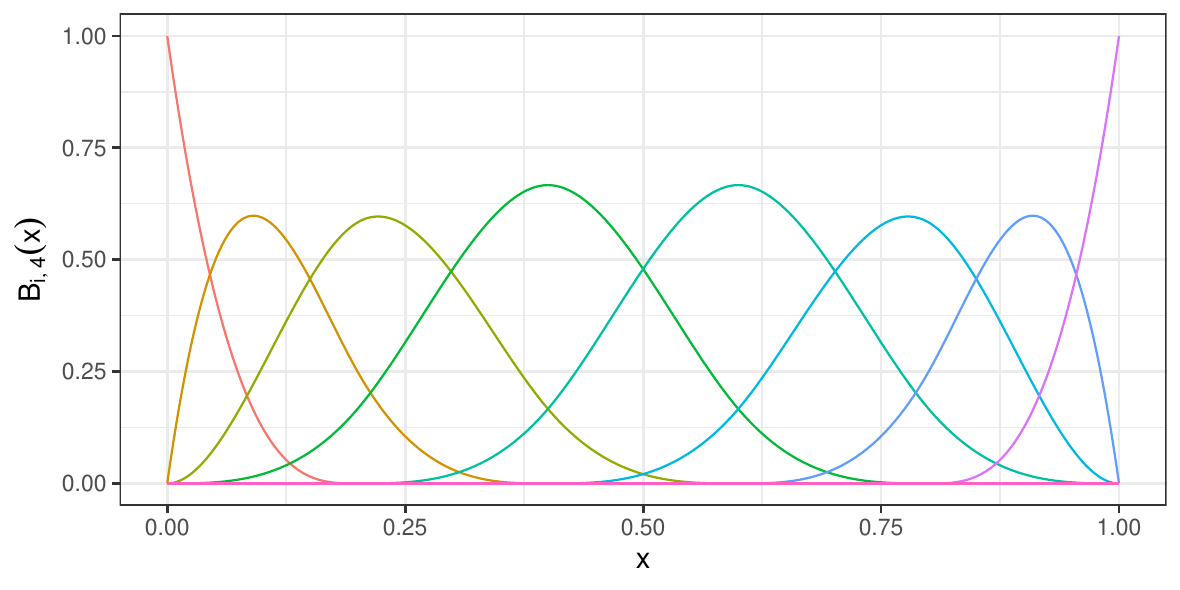}
    \caption{Cubic B-splines with $6$ equally spaced knots ($K=4$) in $[0,1]$.}
    \label{fig: basis splines}
\end{figure}

See Figure \ref{fig: basis splines} for a plot of the cubic (i.e., $M=4$) B-spline basis obtained from $6$ (i.e., $K=4$ internal knots) equally spaced knots in $[0,1]$.
In the software implementation, we used the \texttt{R} function {\texttt bs} from the package {\texttt splines} \citep{bates2024splines, rcoreteam2024r} to evaluate the spline basis functions in the covariates.

\printbibliography

\end{document}